\documentclass[fleqn,usenatbib]{mnras}

\usepackage{newtxtext,newtxmath}

\usepackage[T1]{fontenc}

\DeclareRobustCommand{\VAN}[3]{#2}
\let\VANthebibliography\thebibliography
\def\thebibliography{\DeclareRobustCommand{\VAN}[3]{##3}\VANthebibliography}

\usepackage{diagbox}
\usepackage{graphicx,hyperref,amsmath,mathrsfs,xspace}  
\usepackage{tabularx}
\newcommand{\Flamingo}{\textsc{FLAMINGO}\xspace}
\newcommand{\Mstar}{\mathrm{\Delta M^*_{14}}}
\newcommand{\centshift}{\log_{10}\left<w\right>}

\usepackage[table]{xcolor}

\title[Signatures of dynamical activity in FLAMINGO]{Signatures of dynamical activity in the hot gas profiles of groups and clusters in the FLAMINGO simulations}

\author[]{
Lilia Correa Magnus,$^{1}$\thanks{E-mail: lilia.correamagnus@postgrad.manchester.ac.uk}
Scott T. Kay,$^{1}$
Joop Schaye,$^{2}$
Matthieu Schaller,$^{3,2}$
\\
$^{1}$Jodrell Bank Centre for Astrophysics, School of Physics and Astronomy, The University of Manchester, Manchester M13 9PL, UK\\
$^{2}$Leiden Observatory, Leiden University, PO Box 9513, 2300 RA Leiden, the Netherlands\\
$^{3}$Lorentz Institute for theoretical physics, Leiden University, PO Box 9506, NL-2300 RA, Leiden, The Netherlands
}

\date{Accepted XXX. Received YYY; in original form ZZZ}

\pubyear{\the\year{}}

\begin{document}
\label{firstpage}
\pagerange{\pageref{firstpage}--\pageref{lastpage}}
\maketitle

\begin{abstract}
In anticipation of upcoming cosmological surveys, we use the large volume \Flamingo hydrodynamical simulations to look for signatures of dynamical activity, focusing on the hot gas profiles of groups and clusters out to redshift $z=1$. To determine the dynamical state of each object, we consider the halo mass accretion rate, $\Gamma$, as well as three observational proxies: stellar mass gap, $\Mstar$; X-ray concentration, $c_\mathrm{x}$, and X-ray centroid shift, $\left<w\right>$. In general, the median values of these indicators vary in accordance with an increase in dynamical activity with both mass and redshift. We find $\left<w\right>$ to be the most reliable proxy, while $c_\mathrm{x}$ and $\Mstar$ are more sensitive to resolution and feedback model details. Looking at the profiles, the correlation between dark matter density and $\Gamma$ has a characteristic radial dependence, being negatively (positively) correlated at small (large) radii. This trend is insensitive to both halo mass and redshift. Similar behaviour is also seen for the hot gas densities in low redshift clusters, particularly when using $\left<w\right>$, but the correlations become weaker in groups, at higher redshift and when stronger feedback is employed. We also find the intrinsic scatter in the gas density profiles to decrease with redshift, particularly in groups, contrary to what is seen for the dark matter. Interestingly, the radius of minimum gas density scatter increases with feedback strength, suggesting that this property could be a useful feedback diagnostic in future observational studies. 
\end{abstract}




\section{Introduction}
\label{sec:intro}
Galaxy groups and clusters are of great interest in modern cosmology, they act as tracers of matter on the largest scales. Taking into account the hierarchical nature of large-scale structure formation, the assembly process of these objects leads to embedded characteristic features that act as a record of key events in their history. Clusters have therefore become useful probes of cosmology with their mass and redshift distribution yielding further constraints on the fundamental $\Lambda$CDM parameters (e.g. \citealt{voit2005}; \citealt{Kravtsov2006}; \citealt{Allen_2011}). 

Starting from a simple spherical collapse scenario, \cite{kaiser_1986} predicted that groups and clusters are approximately {\it self-similar} and will, at fixed redshift, have observable properties that follow power-law scalings with mass. Although the reality of large-scale structure evolution is more complex, such power-law relations have been observed (e.g. \citealt{pratt2009,lovisari2020}) and are employed, along with their scatter, for the statistical calibration of cluster masses. 

Departure from the idealized picture is brought about by phenomena that occur outside the monolithic collapse of scale-free dark matter perturbations. Continuous accretion and mergers alter a cluster's dynamical state and perturb the intra-cluster medium (ICM) (e.g. \citealt{Nelson_2012}), causing shocks throughout the gas. Furthermore, feedback from AGN or supernovae is capable of making profound modifications to the ICM around galaxies by injecting large amounts of energy into the system and its progenitors (e.g. \citealt{McNamara2012}; \citealt{Gaspari2014}; \citealt{Voit_2017}). This cumulative disruption throughout a cluster's evolution shifts the location of these objects in the halo property-mass relation, breaking the self-similarity at inner radii in particular (e.g. \citealt{voit2005}; \citealt{McCarthy2017}). 

Both the relaxed and disturbed ends of the cluster dynamical state spectrum present features that aid in classifying clusters and help isolate the more extreme cases in large surveys (e.g. \citealt{raouf_2016}; \citealt{gozaliasl2019}; \citealt{zenteno}; \citealt{yuan2020}). Probing the dynamics involves making use of characteristic properties that can be observed directly. These include observing features in the X-ray emission from the ICM, finding imprints of mergers in the displacement of the CMB spectrum via the Sunyaev-Zeldovich (SZ) effect \citep{SZ_effect} or using the cluster galaxies themselves as probes of assembly history. By taking observables as proxies of dynamical state, classification of clusters into relaxed -- and even cool-core -- or disturbed categories is possible. Finding effective methods utilizing clusters survey data to isolate these extreme regimes with increased precision is an important component of modern cosmology. Characterizing groups and clusters by their dynamical state has been attempted by many previous works, with particular interest in isolating the most relaxed and disturbed cases in large surveys (e.g. \citealt{raouf_2016}; \citealt{gozaliasl2019}; \citealt{zenteno}; \citealt{yuan2020}). This classification involves measuring  characteristic properties of the objects that can be observed directly.  A common optical tracer of merger activity, or lack thereof, is the galaxy magnitude gap. This is the magnitude difference between the brightest cluster galaxy (BCG) and the $N$th brightest galaxy in the system. Since the central galaxy's growth relies primarily on the gas and stars stripped from merged systems, a larger difference or `gap' in magnitude indicates a lack of recent major mergers. Early works focused on its application to studies of dynamically old systems with large magnitude gaps, known as fossil groups (e.g. \citealt{Jones2003}; \citealt{vonBenda_Beckmann2008};  \citealt{dariush2010}). 

In the X-ray, two of the most commonly used probes of dynamical state are the concentration and centroid shift parameters. Concentration, initially introduced by \cite{santos2008} as the ratio of X-ray surface brightness in two apertures; measures the fraction of X-ray emission from within the core, with relaxed (disturbed) objects expected to have large (small) values. Subsequent studies such as \cite{hudson2010}, \cite{rasia_2013} and \cite{Andrade-Santos_2017} have made further use of the X-ray concentration in both simulated and observed cluster data. The centroid shift is a measure of the offset between the X-ray centroid -- taken as the weighted average of X-ray flux -- and the {\it centre} of the cluster (usually the X-ray peak or BCG position). Its sensitivity to the morphology of the gas, with secondary objects shifting the centroid away from the peak, allows the centroid shift to reflect a cluster's current dynamical state (see \citealt{poole2006}; \citealt{maughan_2008}; \citealt{rasia_2013}, \citealt{lovisari2017} and \citealt{Cao_2021} for extensive research on centroid shifts in simulated and observed clusters). 

X-ray data does not only give us a new set of dynamical probes, it also allows us to directly measure the thermodynamic properties of the ICM. By constructing surface brightness profiles and finding the emission measure of these objects, access to projected gas radial densities and therefore gas morphology and distribution is obtained. Spectral fitting of a thermal emission model to the X-ray spectrum is then used to derive temperature profiles (see reviews by \citealt{voit2005} and \citealt{Mroczkowski_2019}). Having obtained these two basic quantities, pressure and entropy profiles are derived assuming an ideal gas and spherical symmetry. Combining dynamical probes and thermodynamic gas profiles results in a variety of possible measurements. For cosmological purposes, such quantities are valuable in producing increasingly accurate X-ray cluster mass estimates (e.g. \citealt{Pointecouteau2005}; \citealt{Arnaud_2010}; \citealt{lovisari2020}). In terms of astrophysical measurements, the imprint left in the ICM by feedback from active galactic nuclei (AGN) can also be probed with these cluster properties (e.g. \citealt{eckert2024}). 

In recent years, large-volume surveys from facilities such as eROSITA \citep{Predehl2021}, the Atacama Cosmology Telescope \citep{Hilton2021}, the South Pole Telescope \citep{Bleem2015} and the Dark Energy Spectroscopic Instrument \citep{Desy2019} have led to a large influx of data —spanning both the X-ray and optical regimes— for cluster cosmology available for cluster cosmology. This trend will continue over the coming decade with \textit{Euclid} \citep{euclid} and the Simons Observatory \citep{simons2019}. These new surveys will be particularly important for finding clusters at higher redshift ($z \approx 1$ and beyond), crucial for improving cosmological parameter constraints and looking for evidence of departures from $\Lambda$CDM.

With these new surveys in mind, we use the state-of-the-art hydrodynamical simulation suite \Flamingo\ \citep{flamingo, kugel2023flamingo}, to study the dynamical states of simulated groups and clusters, and to look for the signatures of dynamical activity (or lack thereof) in the ICM profiles at low and high redshift. Our principal aims are two-fold. First, we aim quantifying how reliably the aforementioned dynamical state diagnostics can identify both relaxed and disturbed clusters in simulations. Secondly, we examine how well measures of dynamical state derived from observational diagnostics are reflected in cluster ICM profiles. The novelty of this work lies in the vast statistical capabilities of \Flamingo, with the benefit of simulation box size allowing robust statistical samples of both groups and clusters to be studied up to $z = 1$.  Additionally, as the suite of simulations was also performed with a variety of feedback models, we also investigate the sensitivity of our results to both (AGN) feedback strength and type (jet vs thermal feedback models).

The rest of the paper is organized as follows. Section \ref{ch:flamingo} provides details of the \Flamingo suite. Our chosen dynamical state proxies are then introduced in Section \ref{ch:Dynamical State Proxies}, along with a description of how we obtained and divided our halo sample. Section \ref{ch:Dynamical State Across Mass Redshift} presents our results on the dynamical probes and their trends with mass and redshift, with Section \ref{ch:thermo profiles} focusing on the thermodynamic profiles of objects with extreme indicator values and how feedback plays a role in what is measured. A final summary of our results and conclusions is provided in Section \ref{ch:conclusion}.

\section{The FLAMINGO Simulations}
\label{ch:flamingo}
We make use of the \Flamingo suite (\citealt{flamingo}; \citealt{kugel2023flamingo}) to perform our statistical analysis of galaxy clusters and groups. This set of large-scale cosmological simulations was run using the SWIFT code \citep{Schaller_2024}; it models cosmology, gravity, and hydrodynamics employing the SPHENIX Smoothed Particle Hydrodynamics (SPH) solver \citep{Borrow_2021}. While detailed descriptions of the simulation setup are available in \cite{flamingo}, the overview provided here covers the most relevant aspects pertinent to our work.

\Flamingo's cosmological setup assumes a spatially flat universe, with parameters adopted from the Dark Energy Survey year three $\Lambda$CDM cosmology \citep{desy_y3}. The subgrid model was built on the foundations developed for the \textsc{OWLS} project \citep{schaye2010} and later used in the \textsc{BAHAMAS} \citep{McCarthy2017} simulations. It incorporates cooling and heating rates from \cite{Ploeckinger2020}, tabulated using the \textsc{cloudy} radiative transfer code (\citealt{cloudy}, version 17.01). Gas is assumed to be in ionization equilibrium, subject to radiation fields such as the Cosmic Microwave Background (CMB), an evolving metagalactic UV/X-ray background (accounting for photoionization), and at high densities, to diffuse interstellar radiation. Cooling temperatures were capped at a lower limit of 100 K; however, such low temperatures are never actually reached in dense gas due to the imposed equation of state which regulates the thermodynamics of the ISM \citep{schaey_dalla_2008}. Finally, the reionization of hydrogen and helium occurs at redshifts $z = 7.8$ and $z = 3.5$, respectively.

Star formation is implemented using the method described in \cite{schaey_dalla_2008}. This involves a stochastic conversion of gas particles into collisionless star particles at a pressure-dependent rate which matches the observed Kennicutt-Schmidt law \citep{Kennicutt1998}. Stellar feedback in the form of winds is implemented using a mass transfer mechanism by which the star particle loses mass to surrounding gas particles (\citealt{wiersma2009}; \citealt{schaye_2015}) whilst supernova feedback is modeled as a kinetic energy injection from core-collapse supernovae (SNe; \citealt{chaikiin_2022}). Finally the seeding, growth and AGN feedback of black holes uses the method presented in \cite{booth2009}, where growth is modeled using a modified version of the Bondi-Hoyle accretion rate. In the \Flamingo Fiducial model, AGN feedback is thermally injected by raising the temperature of the nearest gas particle by $\Delta T_{\mathrm{AGN}}$.

These subgrid prescriptions for feedback were calibrated systematically using Gaussian process emulators (e.g.. \citealt{Rasmussen2006Gaussian}), drawing on data from multiple surveys; the GAMA survey \citep{driver2022} $z=0$ stellar mass function, and the HSC-XXL survey \citep{akino2022} as well as data compiled in \cite{kugel2023flamingo} for gas fractions in galaxy groups and clusters. The maximum halo mass applied to the calibration is resolution dependent (see below). In addition to the \Flamingo \textit{Fiducial} model, eight astrophysical variations were run in a 1 Gpc box, listed in Table 2 of \cite{flamingo}. By altering a total of four subgrid parameters:\{$f_{\mathrm{SN}}$,$\Delta v_{\mathrm{SN}}$, $\Delta T_{\mathrm{AGN}}$, $\beta_{\mathrm{BH}}$\}, these alternative runs were calibrated a given number of $\sigma$ from the observed gas fractions, again applying the emulator method from \cite{kugel2023flamingo}, and are labeled: $\mathrm{fgas}\pm N\sigma$, with $N = \{ \pm2, -4, -8\}$. Another set of runs was produced with a kinetic jet-like AGN feedback prescription defined as the Jet model runs (more details about this model are provided in \citealt{husko_2022}).

\Flamingo was run at three different resolutions, each calibrated to the same datasets but containing varying particle counts depending on the box size and resolution. The flagship runs, as detailed in Table \ref{tab:sim_params}, include box sizes of 1 Gpc and 2.8 Gpc, with high and intermediate resolutions featuring initial gas particle masses of approximately $10^8$ $\mathrm{M_{\odot}}$ (labeled m8) and $10^9$ $\mathrm{M_{\odot}}$ (labeled m9), respectively. We note that all model variations were run using the 1 Gpc box with m9 mass resolution. Each resolution was calibrated separately to different mass ranges for the hot gas fractions, with the m8 simulation using group-scale objects up to $M_\mathrm{500c} = 10^{13.73} \mathrm{M_{\odot}} $\footnote{ $M_\mathrm{500c}$ is the total mass enclosed within a radius $R_\mathrm{500c}$, within which the average interior density is 500 times the critical density of the Universe.} and m9 runs using cluster scale objects up to $M_\mathrm{500c} = 10^{14.36} \mathrm{M_{\odot}}$. As such, parameters for the supernova and AGN feedback subgrid models vary with resolution. \cite{braspenning2024flamingo} have already shown that the Fiducial models have a high level of agreement with observations of gas profiles, which ensures a level of robustness among the \Flamingo group and cluster objects. 

Structures within the snapshots were identified using \texttt{HBT-Herons} \citep{moreno2025}, an improved version of the Hierarchical Bound Tracing algorithm \texttt{HBT+} \citep{han2017}. This history-based approach identifies self-bound objects using a friends-of-friends (FOF) algorithm at each snapshot and incorporates prior outputs to inform the current particle-association step. We also make use of the complementary Spherical Overdensity and Aperture Processor (SOAP\footnote{SOAP is a tool developed as part of the FLAMINGO project. The code is available at \href{https://github.com/SWIFTSIM/SOAP}{https://github.com/SWIFTSIM/SOAP}}) catalogues \citep{McGibbon_2025}, which contain precomputed halo properties across a range of apertures. 

\begin{table}
    \centering
    \caption{Basic simulation details: the label (Name) given to each simulation analysed, the box size ($L$), the number of baryonic particles ($N_{\mathrm{b}}$, which is the same as the number of dark matter particles) and the gas and dark matter particle masses ($m_{\mathrm{g}}$ and $m_{\mathrm{CDM}}$).}
    \label{tab:sim_params}
    \begin{tabular}{lccccl}  
        \hline
        Name & $L$ (cGpc) & $N_{\mathrm{b}}$ & $m_{\mathrm{g}}\ (\mathrm{M_{\odot}})$ & $m_{\mathrm{CDM}}\ (\mathrm{M_{\odot}})$ \\
        \hline
        L1\_m8 & 1 & $3600^3$  & $1.34 \times 10^8$ & $7.06 \times 10^8$ \\
        L1\_m9 & 1 & $1800^3$  & $1.07 \times 10^9$ & $5.65 \times 10^9$ \\
        L2p8\_m9 & 2.8 & $5040^3$ & $1.07 \times 10^9$ & $5.65 \times 10^9$ \\
        \hline
    \end{tabular}
\end{table}

\section{Dynamical State Proxies \& Halo Sample}
\label{ch:Dynamical State Proxies}
\subsection{Dynamical state indicators}
Our work focuses on the set of dynamical state probes briefly described in Section \ref{sec:intro}. We follow past efforts in studying these proxies to assess the types of clusters they may isolate in surveys, by taking advantage of the large statistical samples in the \Flamingo suite. We also make use of a theoretical diagnostic quantity in the form of the mass accretion rate parameter, $\Gamma$, since it directly measures the recent mass assembly rate of dark matter haloes. Below, we provide definitions for each of the chosen indicators. 

\subsubsection{Halo mass accretion rate}
The halo mass accretion rate, $\Gamma$, is the change in total mass over the previous dynamical time \citep{Diemer_2014}, defined as

\begin{equation}
    \Gamma(a_0) \equiv \frac{\Delta \log_{10}(M)}{\Delta \log_{10}(a)} = 
    \frac{\log_{10}(M_1) - \log_{10}(M_0)}{\log_{10}(a_1) - \log_{10}(a_0)},
   \label{eq:gamma}
\end{equation}

 \noindent where  $a_0 = a(t)$ and $a_1 = a(t-t_\mathrm{dyn})$ are the relevant scale factors and $\mathrm{t_{dyn} = \sqrt{\frac{1}{G \rho}}}$ (with $\rho$ being the density of the halo at\footnote{This is the radius within which the average interior density is 200 times the mean density of the Universe.} $R_\mathrm{200m}$) is the dynamical timescale, $M_0$ is the current mass of the halo (also taken within $R_\mathrm{200m}$) and $M_1$ the mass of its progenitor, one dynamical timescale earlier. As such, $\Gamma$ is sensitive to the time resolution of the simulation outputs, since accurate estimates require that snapshots adequately sample the dynamical timescale. Because $\Gamma$ directly reflects the recent mass assembly history of clusters, we can evaluate other observable probes of the dynamical state by using this theoretical quantity as a guide. This is specifically useful in Section \ref{ch:corrs}, where we assess how well such observables reflect a halo population's evolutionary history by measuring their correlation with $\Gamma$.

\subsubsection{Stellar mass gap}

The magnitude gap is a well-established optical measurement of dynamical state but it is not necessarily the best choice when analysing simulation data. Simulations derive galaxy luminosities using models that depend on stellar metallicities and ages which, in turn, rely on subgrid physics. We instead use the stellar mass gap, which, unlike the magnitude gap, is less sensitive to the recent star formation history, and resolution. The stellar mass gap is not unique to the work presented here, although its use in past works is limited. \cite{Deason_2013} extracted this quantity from observations as a proxy for the halo mass gap measured in simulations, while more recently \cite{kim_2024} converted magnitude gaps into stellar mass gaps to facilitate comparison between observations and simulations. Our definition of stellar mass gap varies slightly from the quantity described in both these studies since we simply switched the luminosities in the magnitude gap equation to ($\log_{10}$) stellar mass

\begin{equation}
    \label{eq:smr}
    \Delta \mathrm{M^*_{14}} = -2.5\log_{10}(M^*_\mathrm{sat}/M^*_\mathrm{BCG}),
\end{equation} 

where $\mathrm{M^*_{sat}}$ and $\mathrm{M^*_{BCG}}$ are the stellar masses of the fourth most massive and the most massive (central) galaxy respectively, where masses are computed within a spherical aperture of 50 proper kpc (pkpc). We initially assume the BCG to be the galaxy containing the most bound particle within $R_{\mathrm{500c}}$ and check whether it is the most massive object, if not, we make the most massive satellite our BCG instead. We also confined the search region for the fourth galaxy to be within $\mathrm{R_{500c}}$. Much like the magnitude gap, the stellar mass gap (hereafter $\Mstar$) traces recent merger events or lack thereof, with larger gaps typically indicating more relaxed objects with assembly histories where more of the mass was accreted earlier on. 

\subsubsection{X-ray centroid shift}
Based on the definition presented in \cite{maughan_2012}, the centroid shift, $\langle w \rangle$, measures how much the projected X-ray centroid (taken as the intensity-weighted centre of the X-ray emission) deviates from the X-ray peak luminosity across different aperture sizes. We therefore define $\Delta_i$ as the distance between the centroid and the cluster’s center of potential -- which coincides with the X-ray peak in the majority of our halos and is equal to the position of the most bound particle -- within the $i$-th aperture. We then take the standard deviation for a set of regularly spaced apertures between, $0.15$ and $1.0~R_{500c}$, as

\begin{equation}
\label{eq:centroid}
\langle w \rangle = \frac{1}{R_\mathrm{500c}} \sqrt{\frac{\sum_{i=1}^{N}(\Delta_i - \langle \Delta \rangle )^2}{N -1}},
\end{equation}

\noindent where $N =8$ is the number of apertures and $\langle \Delta \rangle$ is the mean distance over all $N$ apertures. In general, a larger $\langle w \rangle$ indicates a more dynamically disturbed system, such as one undergoing a major merger.

\subsubsection{X-ray luminosity concentration}

The final observable used in our analysis is the X-ray concentration, $c_\mathrm{x}$, defined by \cite{chex_2022} as

\begin{equation}
    \label{eq:conc}
    c_\mathrm{x} = \frac{L_{\mathrm{x}}(0.15R_{\mathrm{500c}})}{L_\mathrm{x}(R_{\mathrm{500c}})},
\end{equation}

\noindent where $L_\mathrm{x}(R)$ is the X-ray luminosity within a projected radius $R$. Compared to the centroid shift and stellar mass gap, the concentration is less sensitive to mergers and may better reflect the strength of cooling and feedback effects. We found that resolution comparisons at $z=0$ presented differences in the amplitude of concentrations. We attribute this to the variations in feedback strengths with resolution and stronger feedback in the m8 run resulting in smaller BCGs and lower average concentrations. 

For both X-ray-based quantities, X-ray luminosity maps are created using the observer-frame soft X-ray band ($0.5-2$ keV) in line with the ROSAT telescope energy range. Luminosities are calculated as described in \cite{braspenning2024flamingo}. These measurements reflect the emission projected along one direction. Very hot and dense particles recently heated by AGN ($< 15$ Myr) were removed from these maps as in \cite{braspenning2024flamingo} and \cite{kay2024relativistic}, although these particles do not affect our measurements in a significant way.

\subsection{Halo mass and redshift selection}
\label{ch:sample selection}

We adopt a substructure-dependent criterion in order to define an  appropriate halo sample in each of the chosen \Flamingo simulations. The haloes are required to contain a minimum of four galaxies, a limit imposed by the stellar mass gap definition. An additional requisite is placed on the galaxies themselves, requiring the fourth most massive object to contain at least 20 star particles. Based on this alone, a halo mass completion limit for the m8 and m9 resolutions is established. To determine this limit, we take a halo mass range that spans group and cluster scales: $13 \leq \log_{10}(M_\mathrm{500c}/\mathrm{M}_{\odot}) \leq 15$, and divide this into 20 logarithmic mass bins. For each bin, the fraction of haloes that fulfills our galaxy and star particle criteria is calculated. A minimum mass limit was then found where this fraction dropped below 100\%, for m9 this corresponds to $\log_{10}(M_\mathrm{500c}/\mathrm{M}_{\odot}) = 14$ and $\log_{10}(M_\mathrm{500c}/\mathrm{M}_{\odot}) = 13.5$ for m8. The upper limit is set to the highest mass within our chosen range. Our haloes are then divided into three coarser logarithmic bins with $\Delta \mathrm{log_{10}} M_\mathrm{500c} = 0.5$. We will henceforth refer to each logarithmic mass bin as: groups ($13.5-14$), small clusters ($14-14.5$) and large clusters ($14.5-15$)

Redshift limits are set by examining the dynamical state distributions (some of which are shown in Fig. \ref{fig:corrs_m8}). For each mass bin, the redshift cut is applied when the number of haloes becomes smaller than 100 haloes per redshift bin, to ensure statistical robustness. Numbers naturally dwindle first for the higher mass objects, leaving the lowest mass bin with the largest redshift range. Table \ref{tab:z_comp} details the mass bins available for each run.

\begin{table}
    \centering
    \caption{The available simulations for each redshifts and mass bins.}
    \label{tab:z_comp}
    \renewcommand{\arraystretch}{1.5}
    \begin{tabular}{lccc}
        \hline
        \phantom{-}$z$ &  & $\log_{10}(M_\mathrm{{500c}}/\mathrm{M}_\odot)$ & \\ 
        \hline 
        & $13.5$--$14$ & $14$--$14.5$ & $14.5$--$15$ \\
        \hline
        1.0 & L1\_m8 & L1\_m8/m9, L2p8\_m9 & \\
        0.5 & L1\_m8 & L1\_m8/m9, L2p8\_m9 & L1\_m8/m9, L2p8\_m9 \\
        0.0 & L1\_m8 & L1\_m8/m9, L2p8\_m9 & L1\_m8/m9, L2p8\_m9 \\
        \hline
    \end{tabular}
\end{table}

\section{Dynamical State Across Mass \& Redshift}
\label{ch:Dynamical State Across Mass Redshift}
\subsection{Relaxed and disturbed states at low redshift}
\label{sec:limits}
\begin{figure}
    \includegraphics[width=\linewidth]{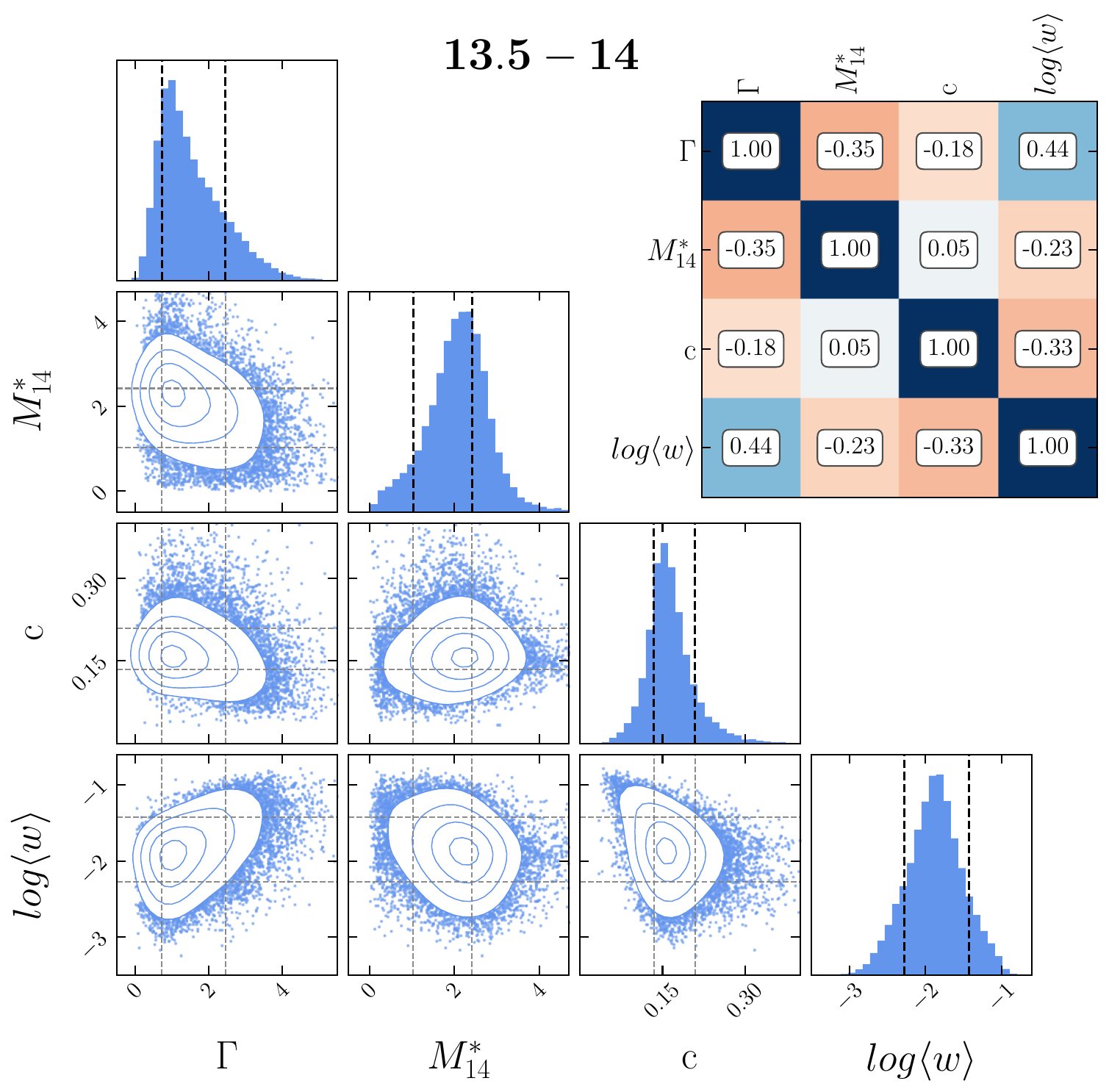} 
    \includegraphics[width=\linewidth]{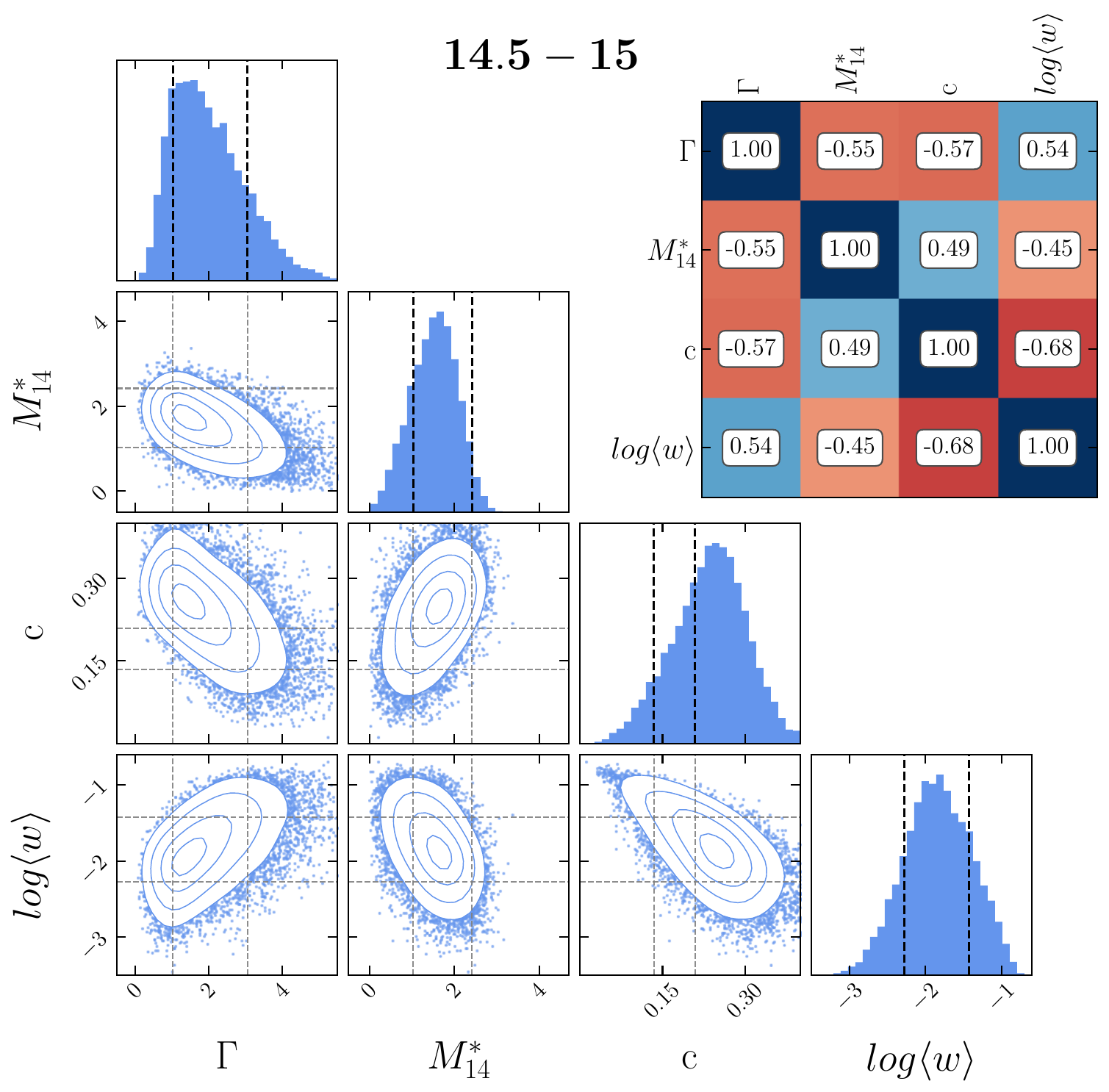} 
    \caption{Corner plots for the L1\_m8 groups: $\log_{10}(\mathrm{M}_\mathrm{500c}/\mathrm{M_{\odot}}) =13.5 - 14$ at $z=0$ (top) and L2p8\_m9 clusters: $\log_{10}(\mathrm{M}_\mathrm{500c}/\mathrm{M_{\odot}}) =14.5 - 15$ at $z=0$ (bottom), comparing the accretion rate, stellar mass gap, X-ray concentration and X-ray centroid shift. In the upper right corners, we present a correlation matrix with the Pearson coefficients for each indicator pair. The dashed vertical and horizontal lines represent the relaxed and disturbed limits for each indicator. Note that the $\Gamma$ limits are calculated for each mass and redshift subset.}
    \label{fig:corrs_m8}
\end{figure}

Similar to the recent study by \cite{towler2023splashback}, we make use of corner plots to visualize the degree of correlation between each dynamical indicator pair for each ($M_\mathrm{500c}$, $z$) sample. Whilst \cite{towler2023splashback} focused on the Fiducial L1\_m9 and L2p8\_m9 simulation, we extend our analysis to the higher resolution m8 run. Fig. \ref{fig:corrs_m8} shows the corner plots for the group [13.5-14] (top) and large cluster [14.5-15] (bottom) samples along with their correlation coefficients at $z = 0$. While both mass bins exhibit qualitatively similar trends the Pearson coefficients are lower for the groups. Such correlation ``wash out'' is likely driven by astrophysical processes such as feedback. We find general agreement with \cite{towler2023splashback} and also with a study that focused on similar dynamical probes by \cite{dariush2010}. They highlight how the magnitude gap alone was unable to locate the majority of the fossils in their sample of simulated galaxy groups. Given our results so far, we can ascribe this problem to the fact that $\Mstar$ -- like our other dynamical state indicators-- exhibits weak correlations in groups. Clusters, on the other hand, typically experience more recent mergers and are less influenced by baryonic processes such as feedback. 

Our X-ray-related probes: concentration and centroid shift, present contrasting results. While their correlations remain relatively strong in high-mass objects, concentration exhibits substantially weaker correlations  than centroid shift in groups. Given this latter quantity is measured using ratios of central and outer apertures around the centres of haloes, AGN feedback affecting the ICM near it could produce an extended hotter gas region, reducing this ratio in groups. 

We use the distributions of our dynamical probes (some of which are shown in Fig. \ref{fig:corrs_m8}) to broadly categorize the halo samples into disturbed and relaxed states. Past efforts have focused on producing ``recipes'' of various observables to accurately isolate the extreme cases in survey samples (e.g. \citealt{deLuca2021}; \citealt{zhang_2022}; \citealt{Casas_2024}). Our principal aim is different, noting that any type of recipe still requires a benchmark to compare with. We instead focus on individual indicators in order to understand whether the relaxed and disturbed objects they characterize have gas properties that reflect these definitions.

We first define values for relaxed and disturbed subsets for each dynamical probe. We refrain from any complex setup and simply take one standard deviation from the mean for a set of low redshift clusters, aiming to get limits that are able to isolate objects in the tail of all dynamical state distributions for each mass and redshift bin. Mirroring the typical mass range of cluster survey samples, we use clusters in the range $\log_{10}(M_\mathrm{500c}/\mathrm{M_{\odot}}) = 14-15$ at $z=0$ and create absolute limits for our observable proxies using the L1\_m8 and L1\_m9 simulations. Table \ref{tab:proxy_lims} shows our limits, defining the relaxed and disturbed subsamples (see also Fig. \ref{fig:corrs_m8}). We note that these do not vary much with resolution. However, we keep limit definitions separate for the m8 and m9 runs due to differences in their subgrid physics implementations. Notably, the m8 run employs an improved black hole repositioning scheme compared to m9 (see \citealt{flamingo}) that can affect the amount of AGN feedback. Whilst the observed probes have fixed limits, we allowed more flexible values for $\Gamma$. We still used the standard deviation from the mean, but calculated the relaxed and disturbed limits for each mass and redshift bin to preserve its role as a theoretical benchmark. 

\begin{table}
    \centering
    \caption{Dynamical state indicator limits for different simulation runs. Each proxy is listed in column one. The next two columns present the relaxed and disturbed limits at m8 resolution whilst the last two present the m9 relaxed and disturbed limits.}
    \label{tab:proxy_lims}
    \begin{tabular}{lcccc}  
        \hline
        Proxy & $\mathrm{m8_{rel}}$ & $\mathrm{m8_{dist}}$ & $\mathrm{m9_{rel}}$ & $\mathrm{m9_{dist}}$ \\ 
        \hline
        $\Mstar$ & >\phantom{--}2.59 & <\phantom{--}1.26 & >\phantom{--}2.39 & <\phantom{--}1.15 \\
        c & >\phantom{--}0.28 & <\phantom{--}0.16 & >\phantom{--}0.29 & <\phantom{--}0.17 \\
        $\mathrm{log\left < w \right >}$ & < -2.32 & > -1.45 & < -2.36 & > -1.47 \\
        \hline
    \end{tabular}
\end{table}

\begin{figure}
\centering
\includegraphics[width=0.9\linewidth]{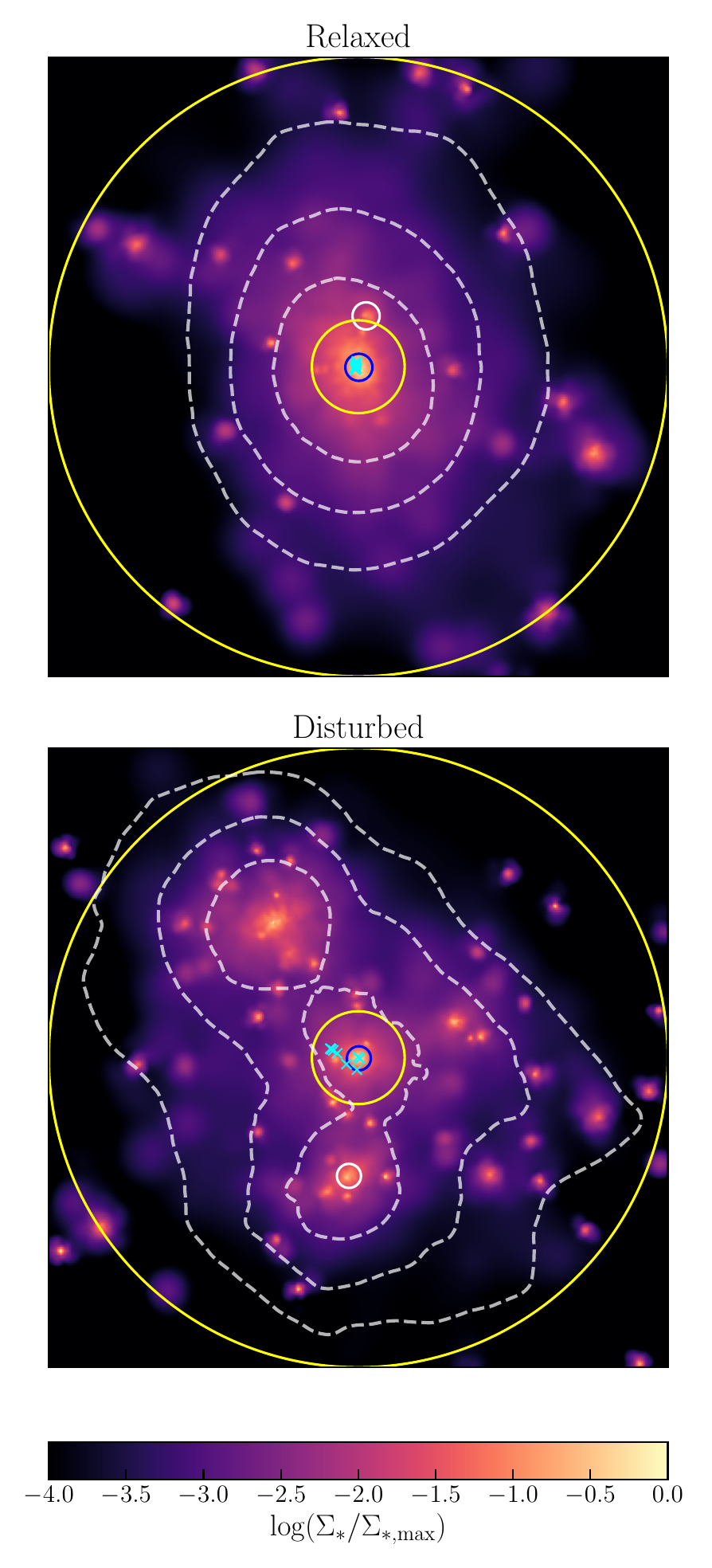}
\caption{Star maps for a relaxed (top) and disturbed (bottom) cluster of similar mass, ($M_\mathrm{500c} \sim 5 \times 10^{14} \mathrm{M}_{\odot}$) at $z=0$ in the Fiducial L1\_m9 run. Highlighted in dashed white lines are the soft band ($0.5 - 2$ keV) X-ray luminosity contours for $\log_{10}(L_\mathrm{x}/L_\mathrm{x,max}) = [-2.5, -1.5, -1.0]$. Both $R_\mathrm{500c}$ and $0.15R_\mathrm{500c}$ apertures are denoted as yellow circles with centroids for the 8 apertures between these radii being marked with cyan crosses. The BCG and 4th brightest object are shown as blue and white 50kpc apertures. The following values are dynamical indicator measurements for each cluster: Merging ($\Gamma = 4.0$, $\Mstar = 0.09$, $c_\mathrm{x}=0.10$, $\log_{10}\left<w\right> = -1.4$) and Relaxed ($\Gamma = 1.5$, $\Mstar = 3.2$, $c_\mathrm{x}=0.29$,  $\log_{10}\left<w\right>= -2.4$).}

\label{fig:maps}
\end{figure}

\begin{figure}
\centering
\includegraphics[scale=0.35]{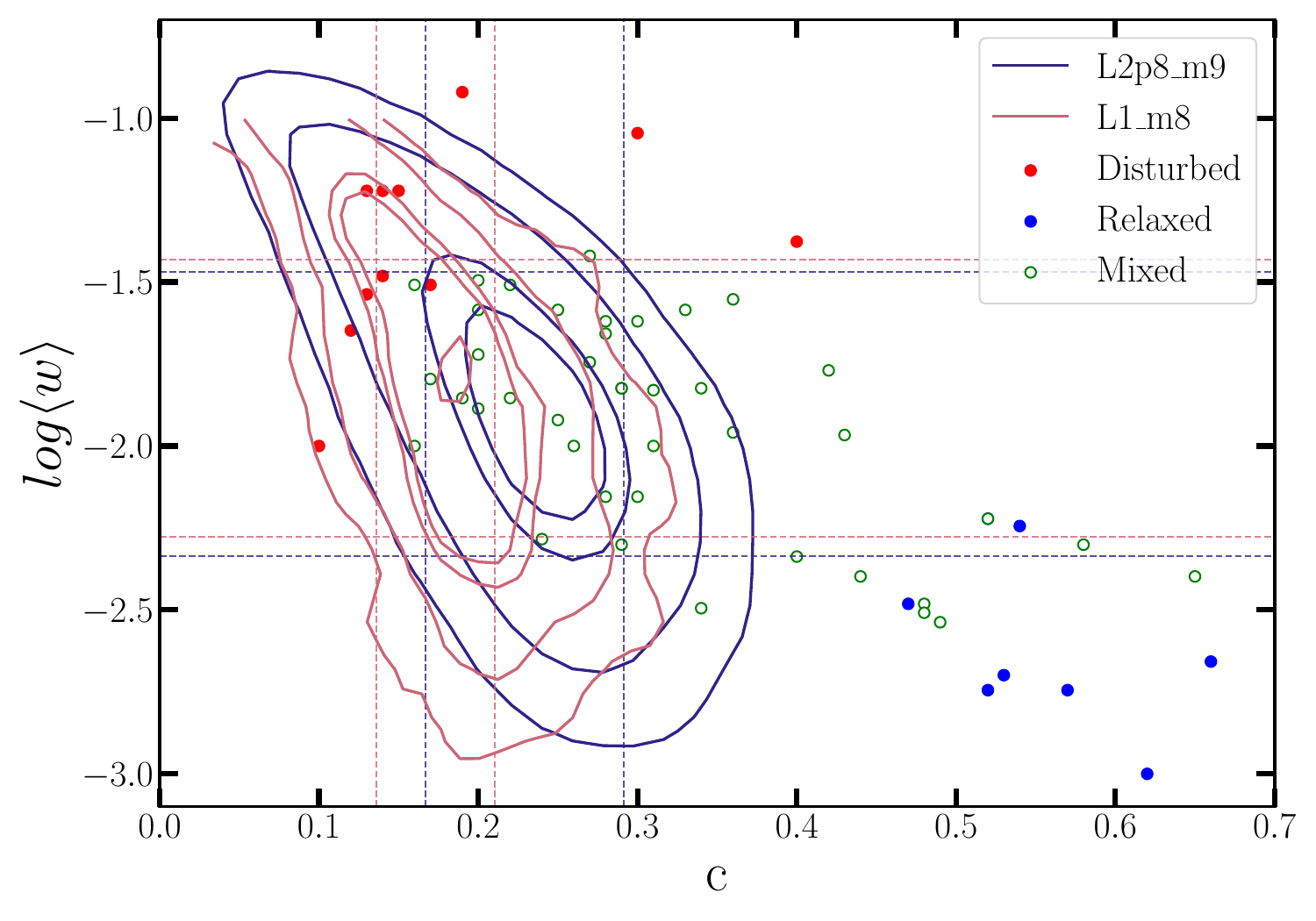}
\includegraphics[scale=0.35]{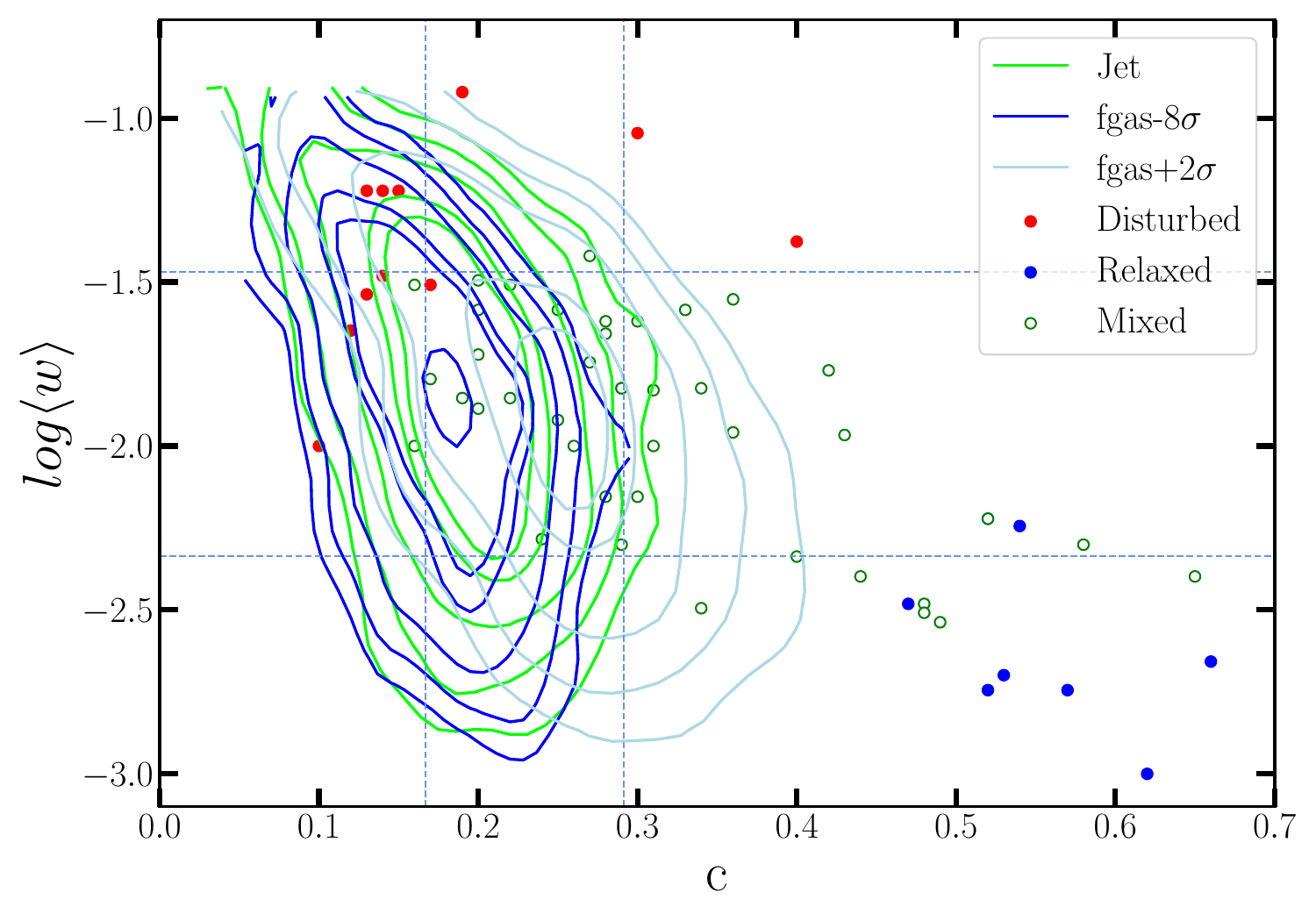}
\caption{\textit{Top}: X-ray centroid shift vs X-ray concentration contour plots for clusters at $z=0$ with $2.5\times10^{14} < \mathrm{M}_\mathrm{500c}/\mathrm{M_{\odot}} < 10^{15}$ in the L1\_m8 and L2p8\_m9 simulations compared to the sample of 61 clusters studied in \protect\cite{chex_2022} in this same mass regime with $z<0.2$. \textit{Bottom}: Same as top but for the Jet,  $\mathrm{fgas}-8\sigma$ and $\mathrm{fgas}+2\sigma$ runs. The dashed lines denote the m8 (orange) and m9 (blue) relaxed and disturbed dynamical limits for the centroid shift and concentration.}
\label{fig:chex_compare}
\end{figure}

Using $\Gamma$ and a `traditional' method of fossil group identification, we are able to locate relaxed and disturbed clusters in the simulations and inspect their dynamical indicator values. We probed clusters of mass: $\log_{10}(M_\mathrm{500c}/\mathrm{M_\odot}) \approx 14.7$ at $z=0$, and sought objects in the tail of the $\Gamma$ distribution ($\Gamma > 4$) to find a merging cluster. For a relaxed system we used the definition of fossils from \cite{Jones2003}, where these are characterized by a magnitude gap $\Delta \mathrm{M^r_{12}} > 2$ and an extended X-ray halo of luminosity $L_\mathrm{x} > 10^{42}\ \mathrm{erg\ s^{-1}}$. To adapt this to our simulation data, we translated the magnitude gap criterion into a stellar mass threshold of $\Mstar > 2.5$, with the slight increase being due to the stellar–mass distribution being skewed toward higher ratios.

Fig. \ref{fig:maps} shows two stellar densities corresponding to clusters of mass $M_\mathrm{500c} \sim 5 \times 10^{14} \mathrm{M}_{\odot}$, chosen using these definitions; clear visual differences between the objects reflect their opposing dynamical states, with the merging cluster having a distorted X-ray halo and multiple large cores that still have not merged. In contrast, a smooth and centrally concentrated X-ray halo is present in the relaxed cluster resulting from the lack of bright secondary objects. Checking the indicator values associated to each object, we find our limits work relatively well for all dynamical probes in the merging cluster, whilst in the relaxed case the centroid shift falls above what we have prescribed. Despite this, there is qualitative consistency with our limits, with our example objects falling at or near the tails of our dynamical probe distribution.

Finally, we compare the X-ray dynamical indicators from \Flamingo with those measured in observed X-ray clusters by \cite{chex_2022}, who focused on low-redshift systems in the \textsc{CHEX-MATE} sample \citep{chex_2021}. Clusters from \textsc{CHEX-MATE} were selected from the \textit{Planck} PSZ2 catalogue, minimizing biases in the sample. We looked at their Tier 1 subset, consisting of 61 low redshift ($0.05<z<0.2$) systems with $2 \times 10^{14} \mathrm{M}_{\odot} < M_\mathrm{500c} < 9 \times 10^{14} \mathrm{M}_{\odot}$. We matched the observed distribution in mass by taking clusters at $z=0$ from our [$14-14.5$] and [$14.5-15$] mass bins for both L1\_m8 and L2p8\_m9 runs, as well as clusters from a set of varying feedback model runs: [Jet, $\mathrm{fgas}-8\sigma$ and $\mathrm{fgas}+2\sigma$].

Fig. \ref{fig:chex_compare} shows the contours in the $\centshift-c_\mathrm{x}$ plane, produced from the \Flamingo clusters and compared to the Tier 1 \textsc{CHEX-MATE} subset (for the latter, we increased the cluster masses by 25 percent to account for hydrostatic mass bias, assuming $b = 0.2$). We see that neither the intermediate (m9) nor high resolution (m8) \Flamingo simulations contain clusters with concentrations as high as those observed, however, the observed range of centroid offsets is more closely reproduced. In terms of the concentrations, the discrepancy is large, with both m8 and m9 runs having mean values of $c_\mathrm{x} \approx0.17$ and $c_\mathrm{x} \approx0.22$ respectively, which fall far below the \textsc{CHEX-MATE} median, $\mathrm{c\approx0.33}$. Our concentration limits (vertical grey lines) therefore compare quite poorly with the real cluster data, having a high contamination of `mixed' observed systems in our relaxed regime. We also study how far our various feedback models are able to change the concentrations. Even the weakest AGN model ($\mathrm{fgas}+2\sigma$) is unable to come close to producing clusters with the highest concentrations, although the median $c_\mathrm{x} -$value in this run ($c_\mathrm{x} \approx0.27$) is closer to that observed in \textsc{CHEX-MATE}. 

In their work, \cite{chex_2022} make a comparison to The300 \citep{The300} simulated clusters, which match the observations relatively closely. Differences between subgrid prescriptions and parameter choices will lead to varying results among simulations. We find that in the case of \Flamingo, lower concentrations are related to the plateau in the cluster entropy profiles \citep{braspenning2024flamingo}. This feature, which departs from observed X-ray cluster profiles, highlights hotter and more diffuse gas found at intermediate cluster radii ($0.1 \lesssim r/\mathrm{R_{500c}\lesssim0.5}$) and lowers the X-ray luminosity concentration. In contrast The300 clusters do not have this distinct characteristic in their entropy profiles \citep{Li2023}, leading to higher concentrations and a closer match to the \textsc{CHEX-MATE} sample.

We should note however, that other observations such as those using the kinetic Sunyaev-Zel'dovich (kSZ) effect favour stronger feedback (e.g. \citealt{mccarthy2025}). This suggests that differences in X-ray luminosity concentrations on their own are not sufficient to justify changes to feedback prescriptions. In further inspecting the cluster sample, we find that $< 1$ percent of the chosen objects values above $c_\mathrm{x}  > 0.5$, highlighting how care should be taken when directly comparing this quantity in our simulations with observations.

\label{ch:Probes Redshift}
\begin{figure}
\centering
\includegraphics[width=0.9\linewidth]{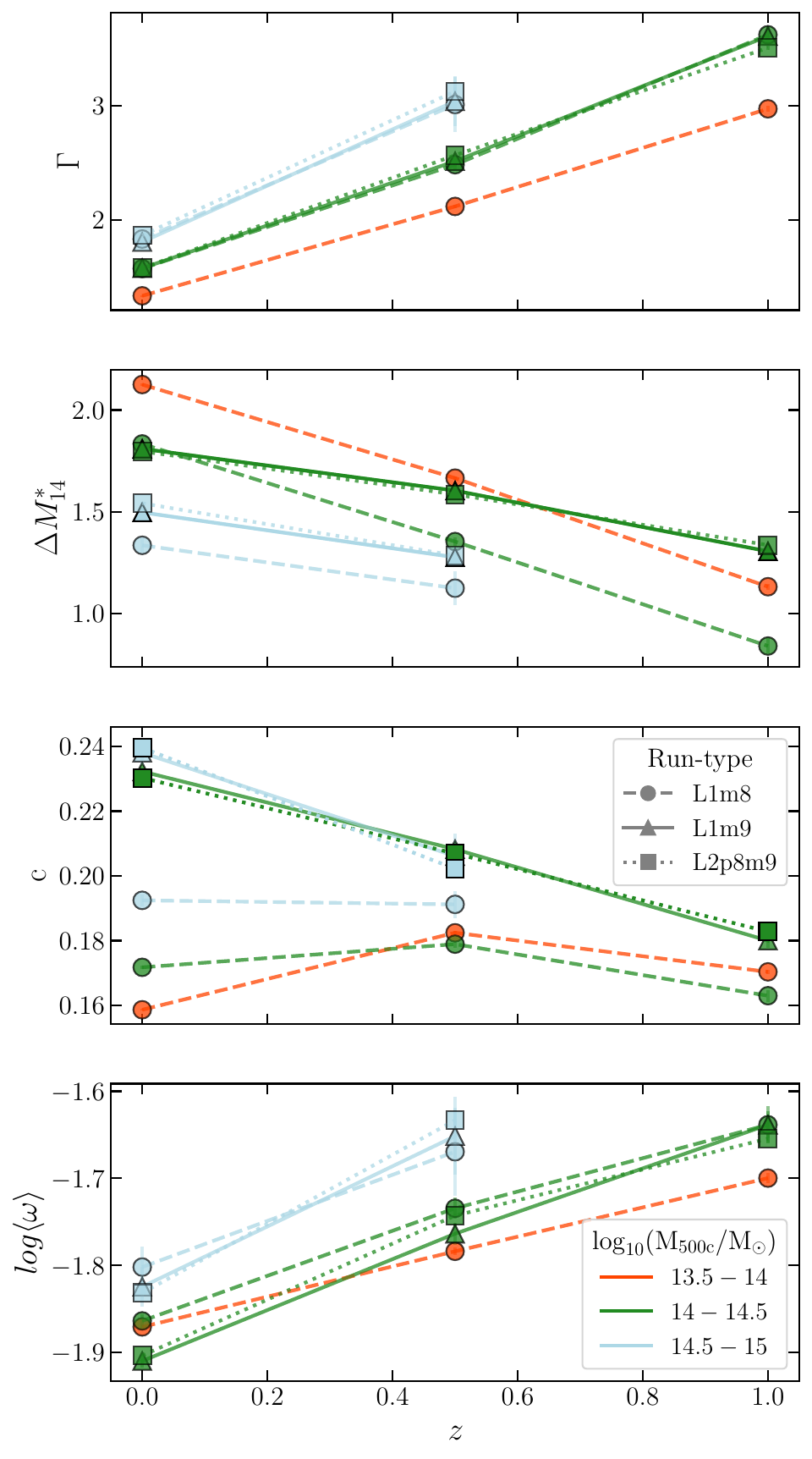}
\caption{Median dynamical state indicator vs redshift for the different mass bins and resolutions for the Fiducial model as listed in Table \ref{tab:sim_params}. The error bars included are the bootstrap standard error of the median.}
\label{fig:medians}
\end{figure}

\begin{figure}
\centering
\includegraphics[width=\linewidth]{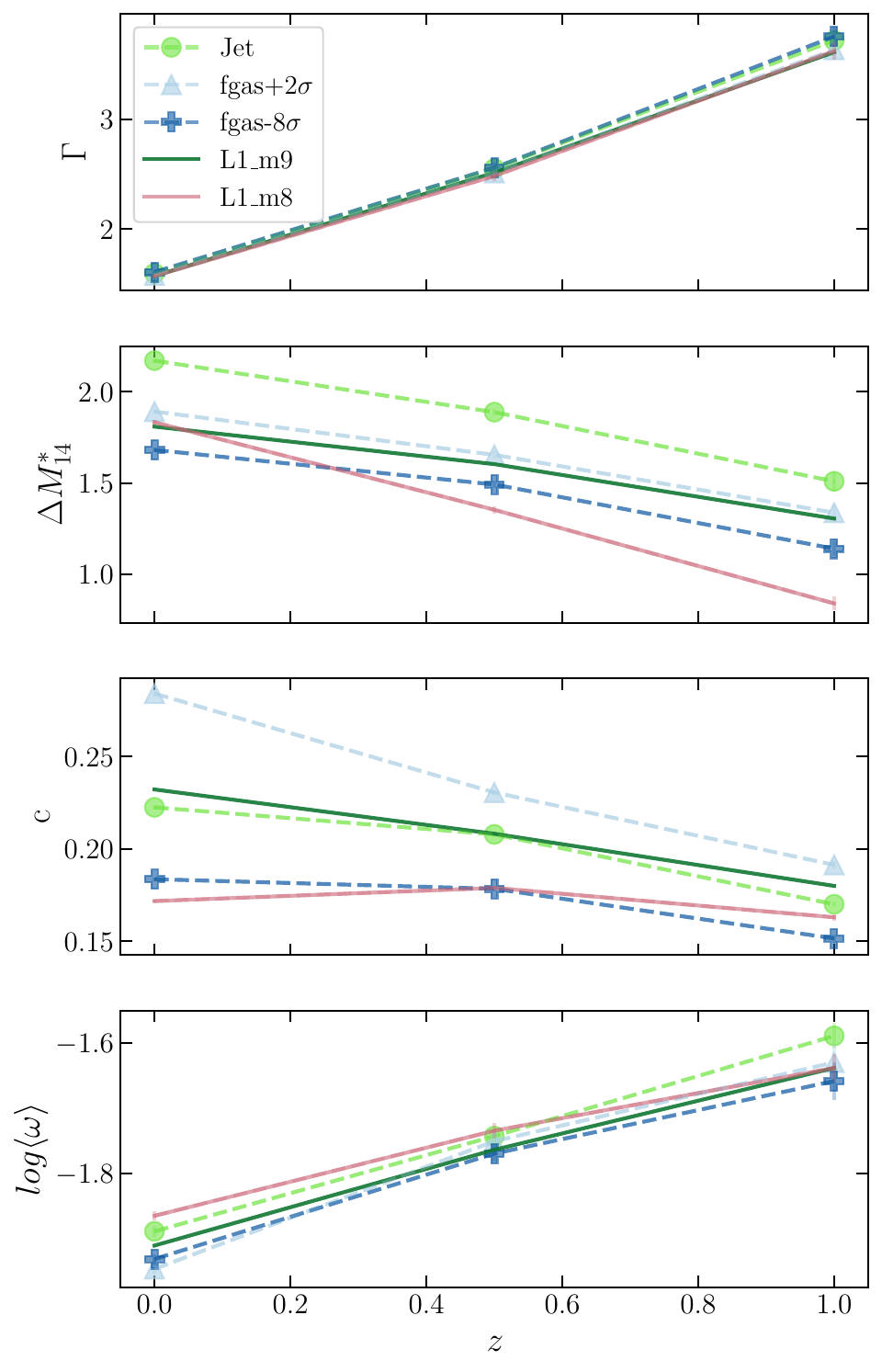}
\caption{Medians of each dynamical state indicator for the $\log_{10}(M_\mathrm{500c}/\mathrm{M_\odot})=14-14.5$ mass bin for varying feedback  runs alongside the Fiducial L1\_m9 and L1\_m8 cases.}
\label{fig:medians_vars}
\end{figure}

\begin{figure}
\centering
\includegraphics[scale=0.5]{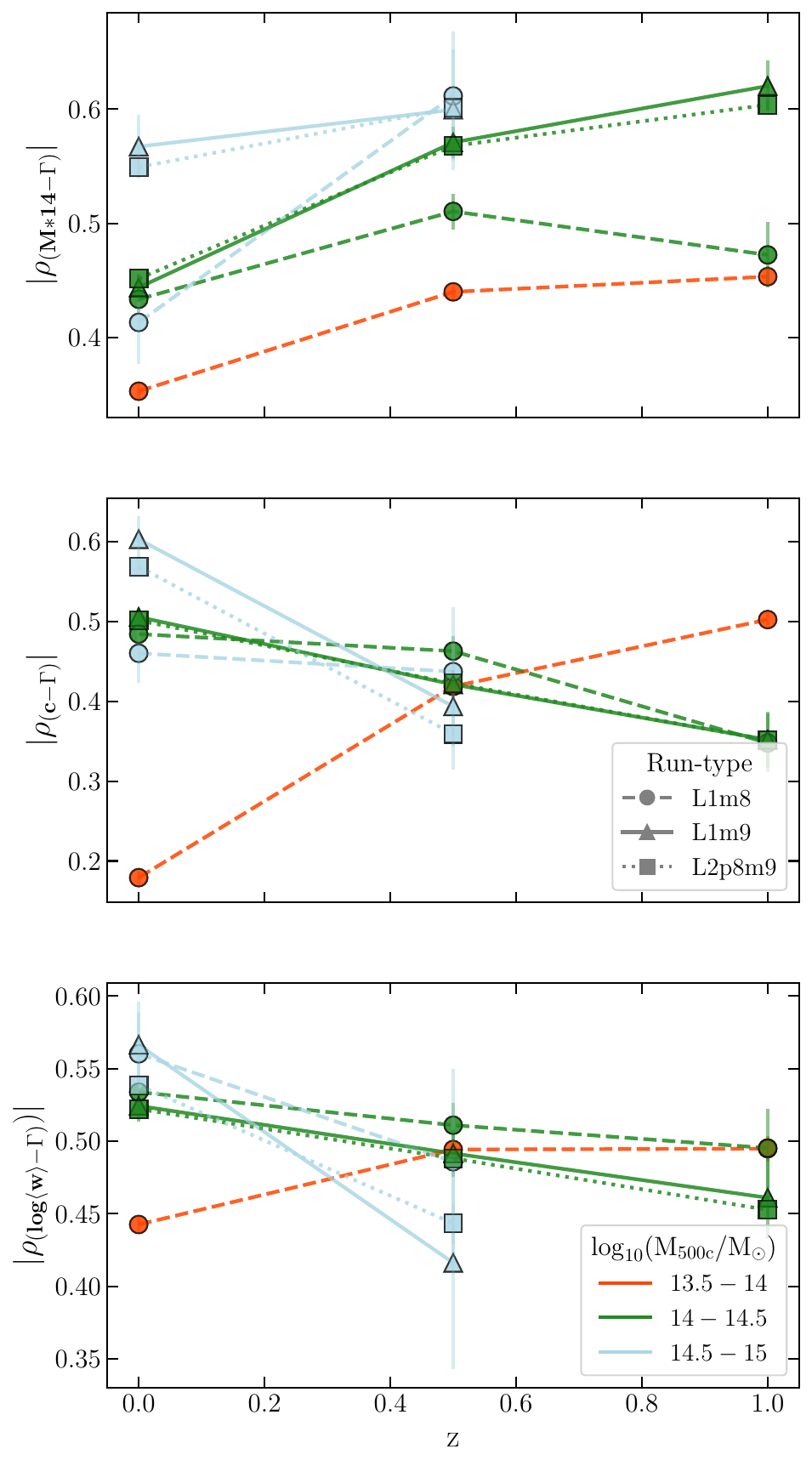}
\caption{Evolution of the absolute Pearson correlation coefficient $\rho$ for each dynamical state indicator with the halo mass accretion rate $\Gamma$, for each ($M_\mathrm{500c}$, $z$) bin available, for the three fiducial simulations. The error bars indicate the bootstrap standard error on the median. }
\label{fig:corrs_redshift}
\end{figure} 

\subsection{Dynamical state indicators across redshifts}
\label{ch:corrs}

We now attempt to understand the redshift dependence of each dynamical state indicator for each of our mass bins. Fig. \ref{fig:medians} presents the redshift evolution of the median indicator value for each ($M_\mathrm{500c}$,$z$) set in the three Fiducial simulations (as defined in Table \ref{tab:z_comp}). For both resolutions we find the indicators to evolve with redshift qualitatively as expected. At fixed redshift, lower mass objects, which tend to form earlier, are typically more dynamically relaxed than the more massive clusters but are also more sensitive to feedback phenomena. We also find, in general, increasing accretion rates and centroid shift values as well as decreasing concentration and stellar mass gaps with increasing redshift, i.e objects at a fixed mass are less relaxed at earlier times.

It is hard to compare our results directly with past observational works, but we can qualitatively assess whether our conclusions match those of observational studies. One such study is presented in \cite{gozaliasl2019}. They focused on the BGG--X-ray peak offset, defining this as the ratio of angular separation between the BGG and X-ray peak to the group's radius: $\Delta r/R_\mathrm{200c}$. Their sample, taken from the COSMOS X-ray survey \citep{Scoville2007}, included galaxy groups with $M_\mathrm{200c} = 8 \times 10^{12} - 3\times10^{14} \mathrm{M}_{\odot}$ at $0.08 < z < 1.53$, being most comparable to our $\log_{10}(M_\mathrm{500c}/\mathrm{M}_{\odot}) = 13.5-14$ groups sample. They found that the BGG offset decreases with increasing mass (at fixed redshift) and decreasing redshift (at fixed mass). Although we use centroid shift, we find this is only true for the offset vs redshift case and see the opposite trend between offset and halo mass. 

We can only speculate about why these results show the inverse trend with mass compared to our simulations and to general theoretical expectation. The observed large offsets in groups could suggest underlying physical processes not fully captured in the models. However, recent work by \cite{Roche_2024} has shown that observational methods using gas can overestimate these offsets by factors of $\sim$30. An additional issue arising from observations is the misalignment between the location of the BGG and the centre of potential in groups. Such objects have been shown to happen in up to $\sim 25\%$ of observed groups \citep{Skibba2010}, potentially creating further departure from our results. We should also note the \cite{gozaliasl2019} results primarily reflect the fact that smaller objects that are merging will have larger offsets. These same systems in our simulations may be ignored or taken to be subhalos within the larger cluster. We are therefore cautious about direct comparisons with such studies without fully taking observational effects into account.

A closer look at each dynamical probe in Fig. \ref{fig:medians} provides insight into both the nature of each measurement as well as features in the simulations themselves. The stellar mass gap vs redshift relation shows resolution-dependent evolution: for the m8 runs, $\Mstar$ has a steeper gradient than in the m9 runs. A smaller $\Mstar$ at earlier times will be caused either by fainter BCGs and/or brighter satellite; we find that fainter BCGs are the dominant driver in our case. However, we should note the most massive clusters in the m8 and m9 simulations were not included in the $z=0$ calibration of the subgrid feedback. Despite matching the data reasonably well, this means that large objects at $z=1$ (the progenitors of the low redshift massive systems), lie outside the calibrated mass range for both resolutions -- see Section \ref{ch:flamingo}. The main property driving this difference is likely to be feedback, with AGN in central BCGs for m8 objects being able to quench star formation more effectively at higher redshift, resulting in the sharper decline in $\Mstar$. 

Looking at the concentration vs redshift relation, we find clear differences between resolutions, noting that the redshift trend in m8 groups has a turnover at $z = 0.5$.  On dynamical state grounds, we would expect $z=0$ groups to have the highest concentrations. However, feedback also plays a role in altering these trends as it couples to the hot gas out to larger radii on group scales. This leads to the hot X-ray halo being more extended, and brings the concentration down. 

Our final observable, the centroid shift, appears to be the most robust probe across resolutions. The m8 and m9 trends agree well, showing larger clusters and systems at higher redshifts to be more disturbed on average. The X-ray peak in simulations is largely determined by the potential minimum, making the centroid shift more resistant to disturbances from feedback. We thus naturally find this to be a good tracer of accretion rate and potentially a good quantity to use when making predictions for observations.

In Fig. \ref{fig:medians_vars} we also look at each dynamical proxy, measured in the $\log_{10}(M_\mathrm{500c}/\mathrm{M}_\odot) = 14-14.5$ logarithmic mass bin, for the most extreme $\mathrm{fgas}\pm\mathrm{N}\sigma$ thermal feedback and Jet feedback model runs. The accretion rates again match well for all models, proving its consistency regardless of feedback parameter changes. Unsurprisingly, we also find that the centroid shift is feedback-model insensitive. For the concentration and $\Mstar$ trends, we see that $\mathrm{fgas}-8\sigma$ (strongest feedback) behaves most similarly to the m8 sample. However, at $z=1$, m8 BCGs are quenched more strongly than in any m9 model. Among the feedback models, variations in the stellar mass of the fourth satellite leads to the observed separation between $\Mstar$ measurements. Although this affects the absolute values of $\Mstar$, the trends are not affected. We also see that, in the concentration measurements, the $\mathrm{fgas}-8\sigma$ and m8 clusters have a similar trend with redshift, flattening at $z=0.5$ and presenting relatively low $z=0$ $c_\mathrm{x}$-values.  Overall, our \Flamingo model variants show that stronger feedback decreases both the stellar mass gap and gas X-ray luminosity concentration, likely due to more effective BCG quenching (in the case of mass gaps) and ejection of low entropy gas (for concentrations).

\begin{figure*}
\centering
\includegraphics[width=\textwidth]{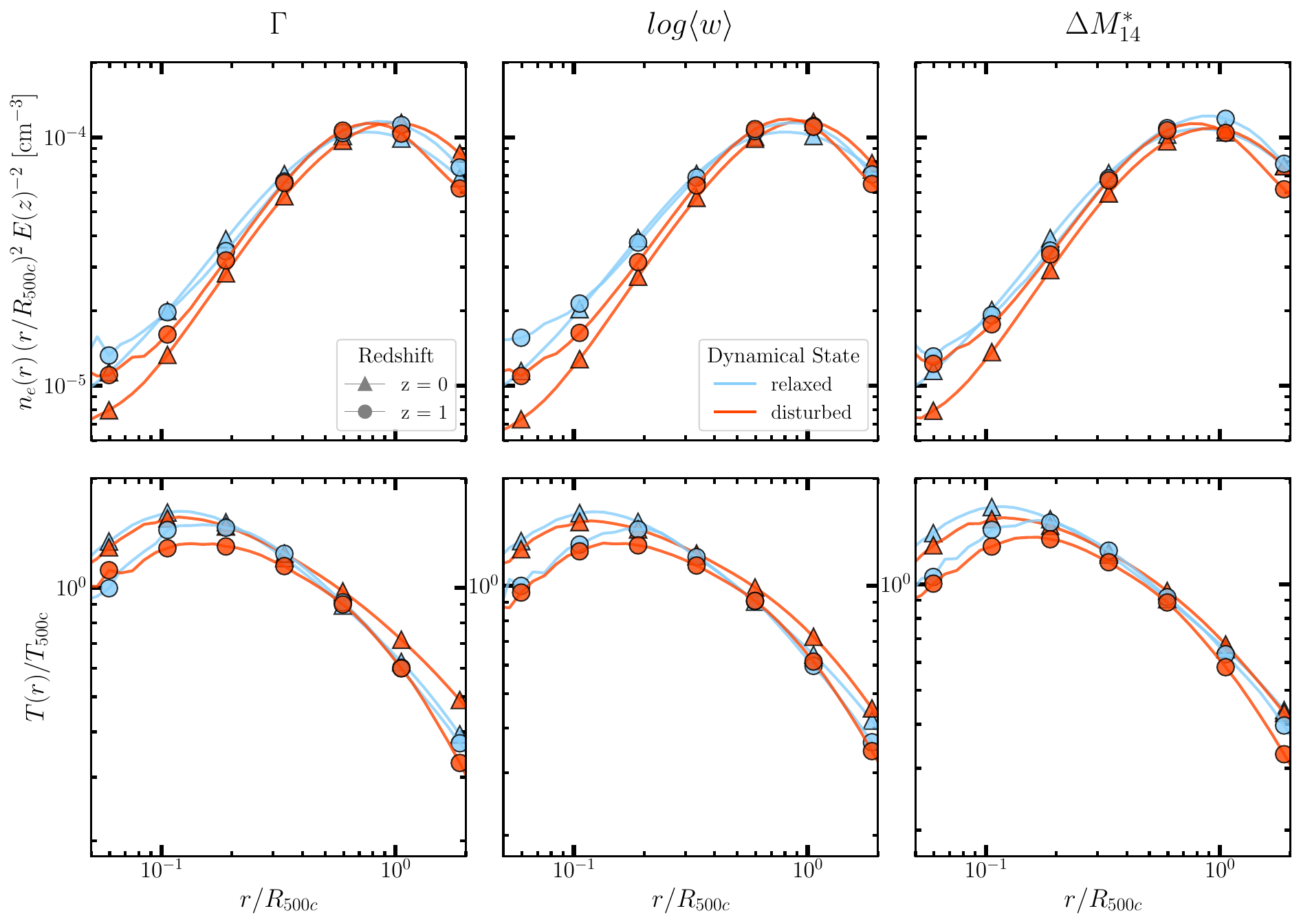}
\caption{Electron number-density (volume-weighted) and temperature (mass-weighted) profiles for disturbed (red) and relaxed (blue) clusters, selected using accretion rate (left), centroid shift (middle) and stellar mass gap (right). Results are shown for the small cluster [$14 - 14.5$] mass bin at $z=0$ (triangles) and $z=1$ (circles) in the L1\_m9 run. }
\label{fig:profiles_dist_rel_gas}
\end{figure*}

We next look at how the correlation strength between the observable proxies and accretion rate $\Gamma$ depend on redshift, for our fixed halo mass bins. Inspection of Fig. \ref{fig:corrs_redshift} shows that the strongest correlations lie at $z=0$ for the X-ray related quantities and at higher redshift for the stellar mass gap. Groups stand out with anomalous trends with redshift, specifically $c_\mathrm{x}$ and $\centshift$ have trends that are opposite from the higher mass objects. Understanding that feedback is most prevalent here, we see how this can also weaken the correlation with accretion rate. The most dramatic drop in correlation strength is observed with groups at $z=0$ in the concentration and centroid shift correlations, with $c_\mathrm{x}-\Gamma$ decreasing to $\rho\sim0.1$ at $z=0$.

\section{Thermodynamic Profiles}
\label{ch:thermo profiles}

\begin{figure}
\centering
\includegraphics[scale=0.45]{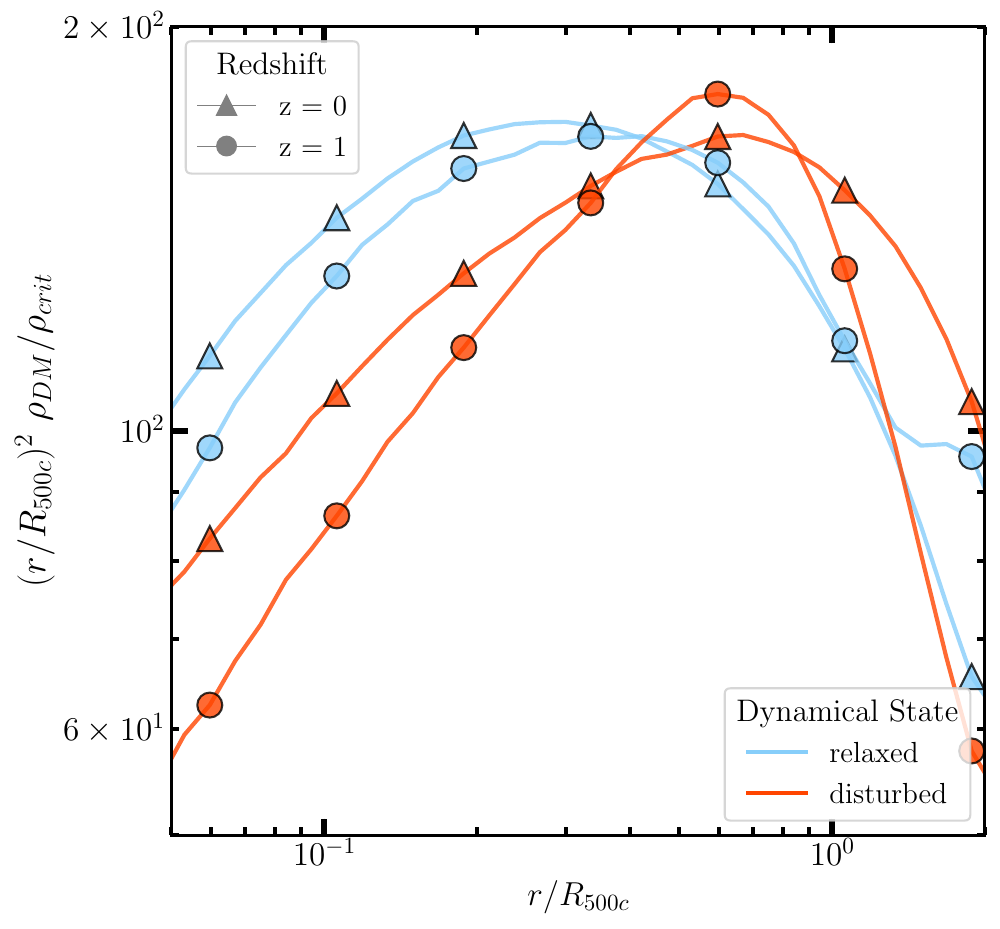}
\caption{Dark matter density profiles for the small cluster [$14 - 14.5$] mass bin at $z=0$ (triangles) and $z=1$ (circles), in the L1\_m9 Fiducial run. Relaxed and disturbed subsamples are defined using $\Gamma$.}
\label{fig:self_sim_DM}
\end{figure}

Our assessment of common dynamical state indicators so far has focused on their mass and redshift trends as well as the strength of their correlations. These indicators are often used in observational studies to distinguish merging clusters from relaxed ones, typically by applying thresholds to their values. We now assess whether the dynamical state predicted by these indicators is reflected in the thermodynamic gas profiles of our groups and clusters. Given that at higher redshifts a larger fraction of systems are disturbed, we aim to take advantage of the redshift range in our \Flamingo runs to find whether we see changes in the profiles at earlier times.

\begin{figure*}
\centering
\includegraphics[width=\textwidth]{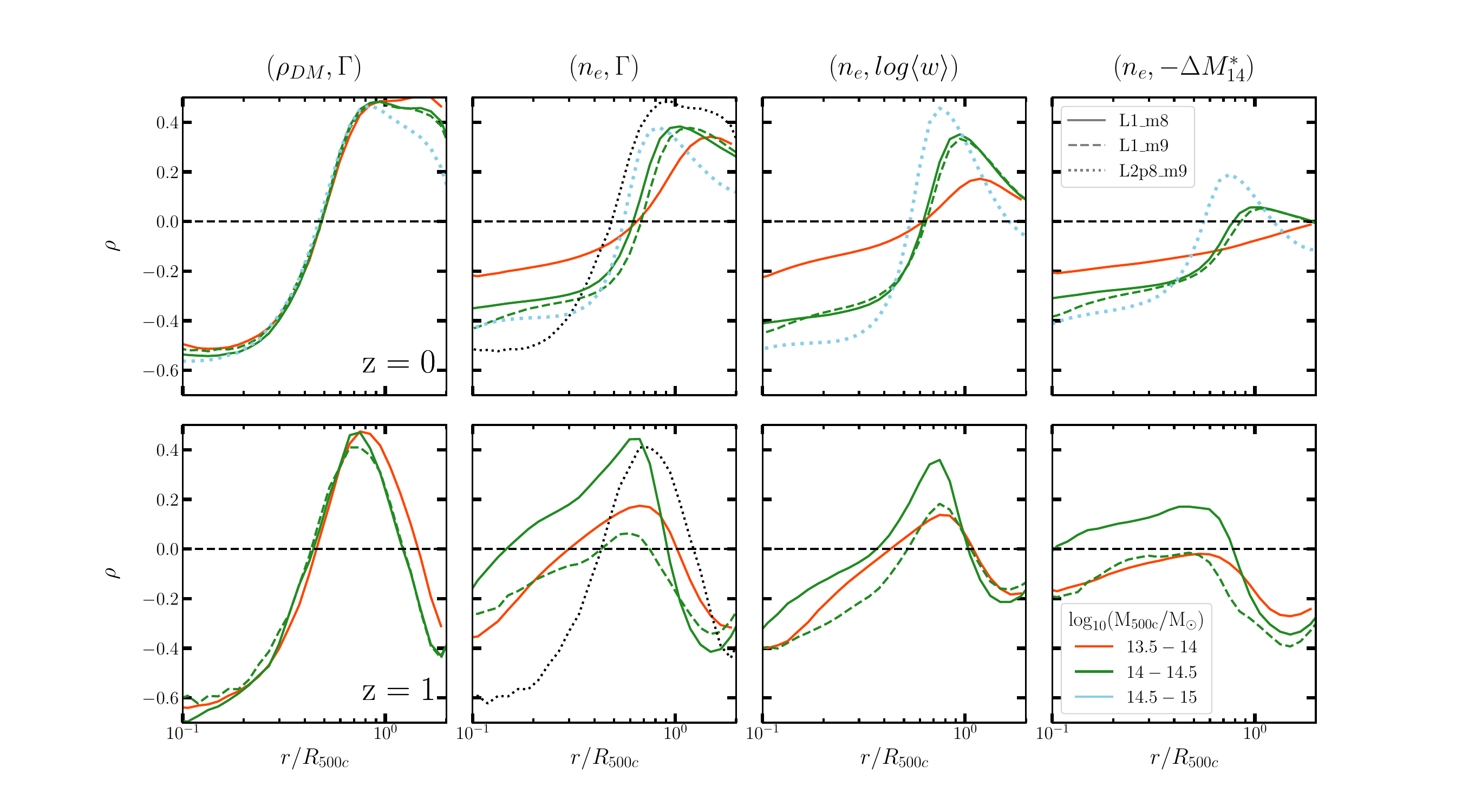}
\caption{Pearson correlation coefficients ($\rho$) for: dark matter density vs $\Gamma$ (first column), and electron density vs $\Gamma$, centroid shift, and stellar mass gap (columns $2-4$). Results are shown at $z=0$ (top row) and $z=1$ (bottom row), including all available halo mass bins at each redshift.}
\label{fig:riva_full}
\end{figure*}

Following \cite{kay2024relativistic} and \cite{braspenning2024flamingo}, electron number density and temperature profiles were calculated as discrete summations of all hot gas ($T > 10^6$K) particles within spherical shells. Each particle's volume was defined as $m_i/\rho_i$, where $m_i$ is particle $i$’s mass and $\rho_i$ its SPH density. Particles recently heated by AGN (within the past 15 Myr) were excluded because the sudden increase in energy makes them briefly artificially hot for their density. The general form of the thermodynamic profile averages is 

 \begin{equation}
    \label{eq:prof}
     \langle A \rangle = \frac{1}{W} \sum^N_{i=1} w_i A_i (m_i/\rho_i),
 \end{equation}

\noindent where $w_i$ is the weight of each particle, $W = \sum^N_\mathrm{i=1} w_i $ the normalisation constant and $A_i$ the property being averaged. For  electron density profiles, we set $w_i = 1$ (volume weighting), which reflects the spatial distribution of electrons. For temperature, we take $w_\mathrm{i} = \rho_i$ (mass weighting), emphasizing the thermal state of the gas by accounting for the mass of each particle. 

Temperatures are presented relative to the self-similar scaling

\begin{equation}
    \label{eq:T500}
    T_\mathrm{500c} = \frac{GM_\mathrm{500c}\mu_\mathrm{e} m_\mathrm{p}}{2k_\mathrm{B}R_\mathrm{500c}} ,
\end{equation}

\noindent where the mean atomic weight per electron is taken as $\mu_\mathrm{e}=1.14$, whereas densities are scaled by $\mathrm{E}(z)^2 \propto \rho_\mathrm{crit}$.

\subsection{Median profiles of relaxed and disturbed objects}

We apply the limits from Table \ref{tab:proxy_lims} to split the ($M_\mathrm{500c}$, $z$) samples into disturbed and relaxed objects, and plot their median gas profiles at $z=0$ and $z=1$ in Fig. \ref{fig:profiles_dist_rel_gas}. Our analysis is limited to $\Mstar$ and $\centshift$ given that our prior results showed X-ray luminosity concentrations to be a poor dynamical state proxy in the \Flamingo simulations. We also use the accretion rate, $\Gamma$, and again make this our benchmark probe, taking $\pm1\sigma$ for each ($M_\mathrm{500c}$, $z$) bin to define the limits (as explained previously). 

The left-most column (using $\Gamma$ limits) in Fig. \ref{fig:profiles_dist_rel_gas} shows how relaxed (blue) clusters have a high degree of self-similarity with redshift, with temperature profiles changing only in the core. Disturbed objects (red), on the other hand, show relatively larger differences between $z=0$ and $z=1$. Compared to the $z = 0$ sample, merging objects at $z=1$ have higher densities and lower temperatures than their low redshift counterparts. This is an initially puzzling result since a larger number of recent mergers (which occur at higher $z$) might be expected to increase the temperature of the gas and lower its density on average. We find, however, that the disturbed objects for all three indicators show a similar result.

To check whether these differences are gravitational in origin, we plot the dark matter density profiles in Fig. \ref{fig:self_sim_DM} for the same L1\_m9 Fiducial run clusters (with disturbed and relaxed objects selected using $\Gamma$). We find a clear separation between relaxed and disturbed objects at both redshifts, something we did not see in the gas. Relaxed objects have higher dark matter densities at small radii ($r<0.5R_\mathrm{500c}$) and the $z=1$ objects move towards lower densities in both cases. Since we are selecting haloes with the same mass at the two redshifts, the disturbed systems have more mass at larger radii, reflecting the mergers with large secondary objects. Our results suggest that astrophysical processes (cooling in particular) must be enhancing the gas density in the merging clusters at high redshift. 

\subsection{Correlation between gas profiles and dynamical state}
\label{ch:Correlations}
\begin{figure}
\centering
\includegraphics[scale=0.6]{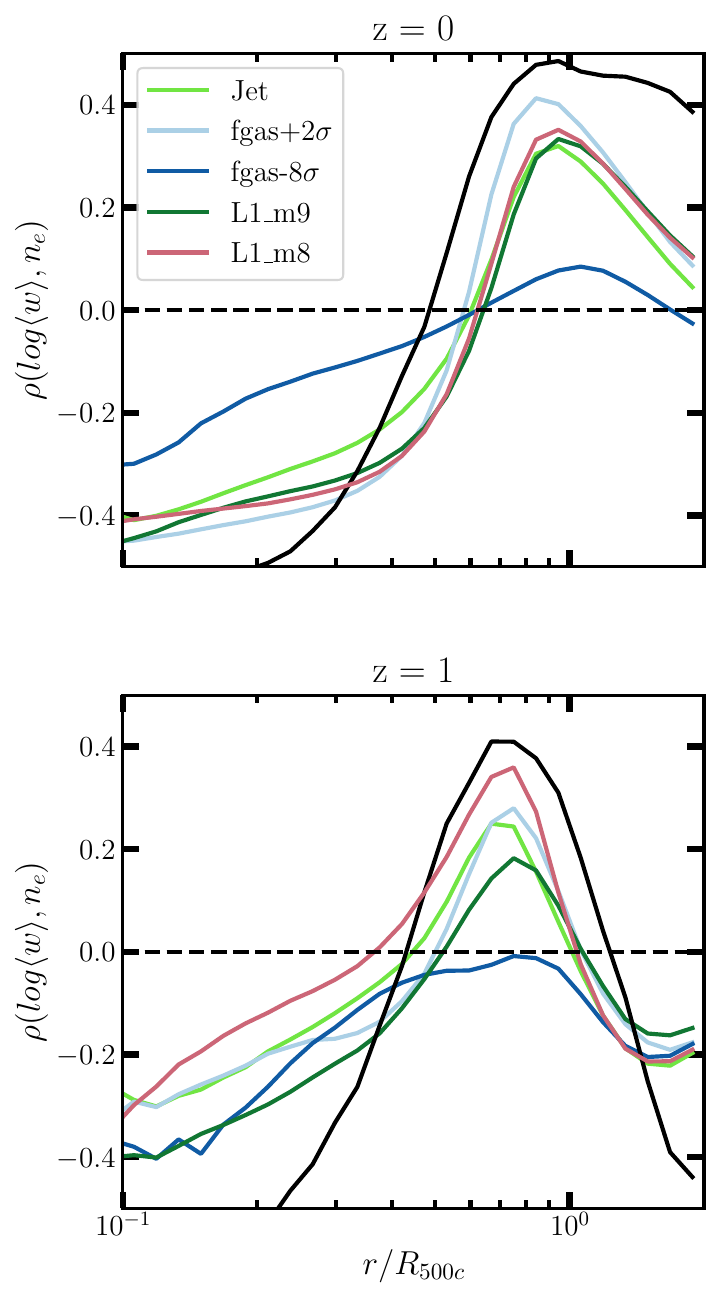}
\caption{Pearson $\rho$ for the $\centshift - n_e$ relation for clusters of mass [$14-14.5$] at redshifts $z=0$ (top) and $z=1$ (bottom) in the most extreme fgas \Flamingo models, the Jet run and the L1\_m9 and L1\_m8 Fiducial simulations. As in Fig. \ref{fig:riva_full}, the $\rho(\Gamma,\ \rho_\mathrm{DM}/\rho_\mathrm{crit})$ curve for L1\_m9 clusters is shown in black for qualitative comparison.}
\label{fig:riva_model}
\end{figure}

We now specifically look at the correlations between the thermodynamic quantities and our dynamical state indicators at each radius. As in \cite{Riva2024}, we calculate the Pearson’s $\rho$ between a gas property and our three main dynamical probes ($\Gamma$, $\Mstar$ and $\centshift$). We do this by dividing each halo into spherical shells from the center out to $R_\mathrm{500c}$ and compute the median correlations in each radial bin for the different halo subsets. Here we focus on the density profile to make direct comparisons between the gas and dark matter components. \cite{Riva2024} used CHEX-MATE data as well as the MACSIS \citep{macsis} and The300 \citep{The300} simulated clusters. However, their results are limited to the most massive objects, defined as the HIGHMz subset with masses $\log_{10}(M_\mathrm{500c}/\mathrm{M}_{\odot})\sim 15$ at redshift $\sim 0.3$. Fig. \ref{fig:riva_full} shows our results for dark matter density and $\Gamma$, as well as the electron number density correlation with $\Gamma$, $\Mstar$ and $\centshift$, for groups and clusters in the fiducial simulations.

Comparing dark matter and gas density correlations with accretion rate (first two left columns of Fig. \ref{fig:riva_full}), we find gas will correlate differently depending on mass and redshift. In contrast, the dark matter component retains a very similar trend within $R_\mathrm{500c}$ for all objects at both redshifts. At higher redshift, astrophysical processes seem to wash out the gas density correlation with accretion rate in groups and small clusters at small radii ($r < 0.3 R_\mathrm{500c}$). We also see how, unlike in the dark matter case, the radius at which $\rho$ changes sign decreases at higher redshift for fixed mass. Analyzing the $\Mstar$ and $\centshift$ correlation curves ($3^\mathrm{rd}$ and $4^\mathrm{th}$ columns in Fig. \ref{fig:riva_full}) we find the centroid shift case resembles the accretion rate trend both at low and high redshift. The stellar mass gap, however, shows weaker correlations across both mass and redshift.

To further probe the effect of feedback, we choose one of our indicators and see how its correlation with the gas density changes with feedback strength. Fig. \ref{fig:riva_model} shows the $\centshift - n_e$ Pearson coefficients for the small cluster sample [$14-14.5$] in the same \Flamingo simulations selected in Section \ref{ch:Probes Redshift}. Focusing first on the amplitude of the curves, we find feedback strength can alter the correlations across radii. It is particularly interesting to see that the strongest thermal model,  $\mathrm{fgas} - 8\sigma$, remains the largest outlier in the sample, with weaker correlations both at low and high redshift. The L1\_m8 simulation shows the weakest correlation at higher redshift in the central regions, $r < 0.3\ r/\mathrm{R_{500c}}$.

\subsection{Profile Scatter}
\begin{figure}
\centering
\includegraphics[width=\linewidth]{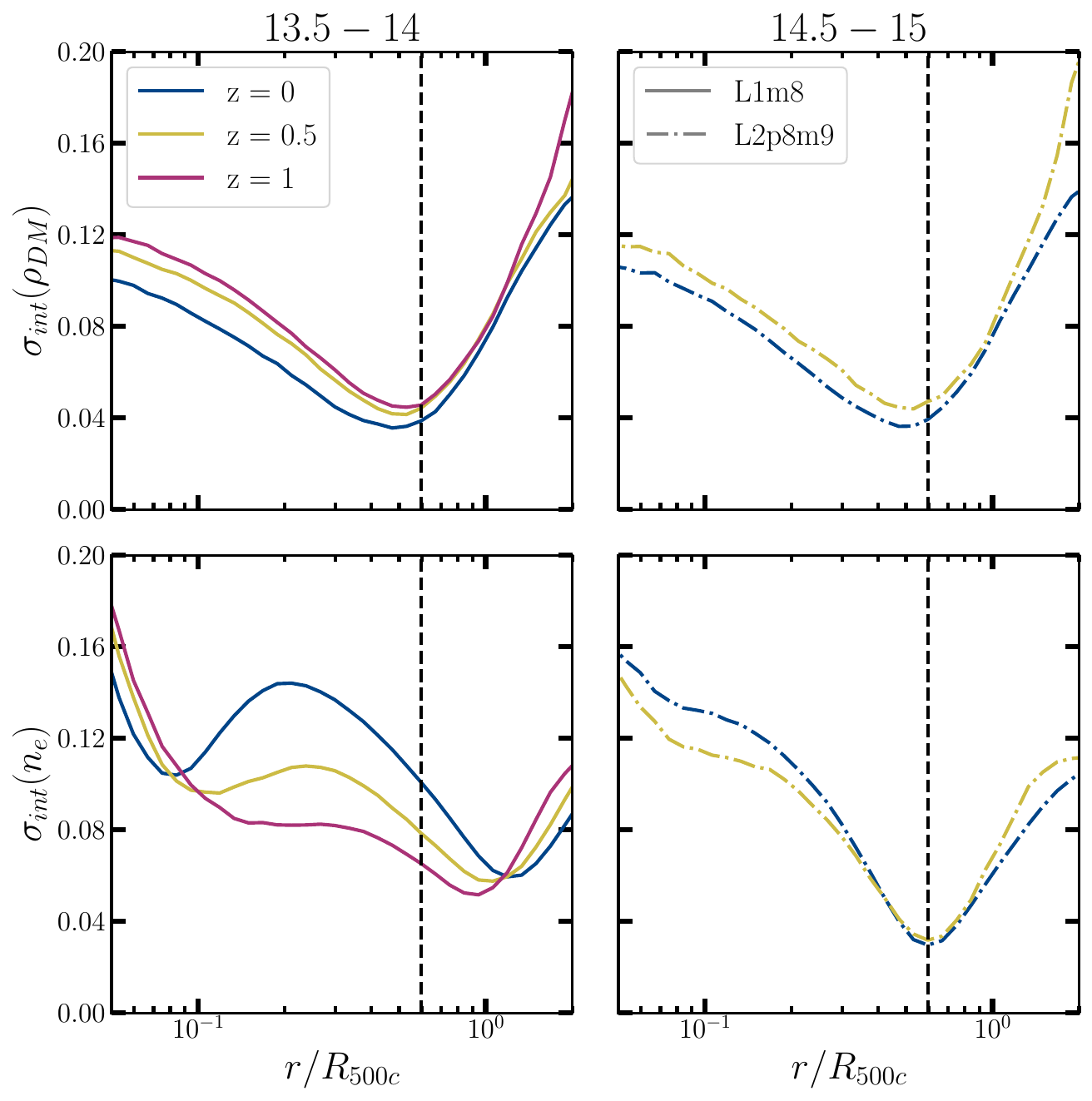}
\caption{Intrinsic scatter in the dark matter density (top) and electron number density (bottom) profiles for groups (left) and large clusters (right). The dashed vertical lines indicate the radius of the global minimum in the dark matter scatter at $z=0$.}
\label{fig:scatter_z}
\end{figure}

\begin{figure}
\centering
\includegraphics[width=0.8\linewidth]{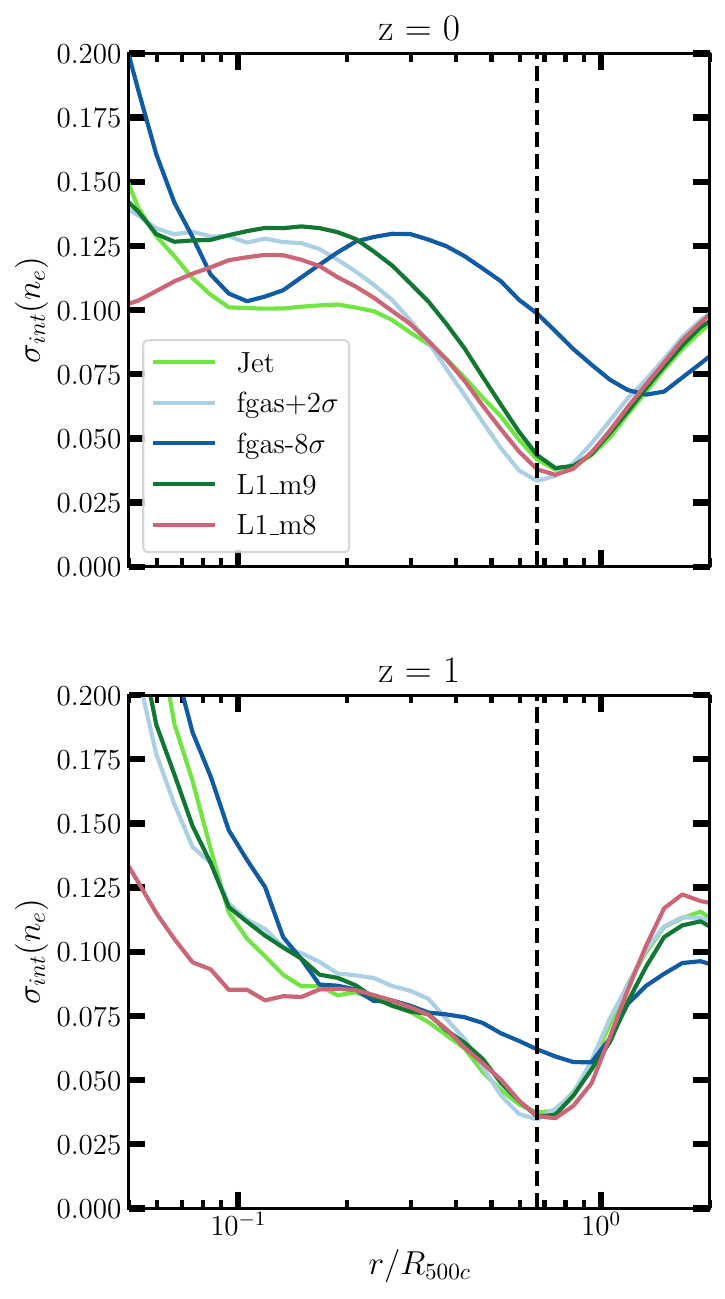}
\caption{Intrinsic scatter in the electron number density profiles in the varying models for the $14-14.5$ clusters at $z=0$ and $z=1$. The dashed vertical lines indicate the radius of the minimum for the intrinsic scatter in the dark matter profiles at $z=0$.}
\label{fig:scatter_model}
\end{figure}

We can also test how feedback is affecting cluster thermodynamic gas properties by looking at how the intrinsic scatter of the profiles varies with mass and redshift. The dispersion will be larger when the profiles for clusters are more affected by mergers and should result in dynamical indicators correlating well with profile measurements. We have seen that on cluster scales, our indicators have relatively high Pearson coefficients but these decrease when feedback has a stronger influence (e.g. in smaller objects at fixed redshift). 

Fig. \ref{fig:scatter_z} shows the intrinsic scatter, $\sigma_\mathrm{int}$; defined as the difference between the $84^\mathrm{th}$ and $16^\mathrm{th}$ percentiles of the log-density profiles, for groups ($13.5 - 14$, L1\_m8) and large clusters ($14.5-15$, L2p8\_m9) in our Fiducial runs. As expected, we find the scatter in the dark matter density (top) to increase with redshift in both groups and clusters, reflecting these systems being more disturbed at higher redshift. The radius of minimum scatter remains unchanged with mass and redshift in the dark matter component but evolves with mass in the gas quantities. We find that the scatter in the gas density in high-mass cluster objects has a minimum that aligns well with the dark matter minimum. In contrast, group-scale objects have minimum locations shifted towards larger radii. This shows the increasing influence of feedback in such systems. Within this ``feedback radius of influence'', the scatter in gas density decreases with increasing redshift, contrary to the dark matter, reinforcing the idea that feedback weakens the dynamical state signal at higher redshift. These results are consistent with \citealt{Lucie-Smith2025}, who found that feedback is most efficient at group scales. They reported this efficiency to be largely independent of redshift at fixed halo mass. However, we find a small redshift dependence; the largest shifts in the scatter minimum are driven by changes in mass scale due to feedback becoming less effective in clusters than in groups.

We also test how feedback affects the scatter by measuring $\sigma_{int}$ for our different feedback models, expecting the minimum to shift with different feedback strengths. Fig. \ref{fig:scatter_model} shows the scatter in the electron density profiles for clusters in the $14-14.5$ logarithmic mass bin at $z=0$ in our different model runs. There is a shift in location of the minimum with feedback strength in the gas density scatter whilst $\sigma_\mathrm{int}$ for the dark matter component remains unchanged, both with model and redshift. In particular, the $\mathrm{fgas} - 8\sigma$ run (featuring the strongest feedback) has a minimum above $R_\mathrm{500c}$ and more scatter outside the core whilst our weakest feedback model ($\mathrm{fgas} + 2\sigma$) has a minimum close to the dark matter, at $\sim 0.7 R_\mathrm{500c}$. 

We can make qualitative comparisons to the measured scatter in observed cluster profiles, like those from CHEX-MATE, REXCESS \citep{rexcess}; X-COP \citep{Xcop} and ESZ \citep{planck2011}, \cite{Riva2024} and \cite{Rossetti_2024} studied the intrinsic scatter in the entropy and projected temperature profiles respectively, and measured these for clusters in the aforementioned surveys, with their samples being most similar to our $\log_{10}(M_\mathrm{500c}/M_{\odot}) =14.5-15$, $z=0$ sample. \cite{Riva2024} find the observed X-ray clusters to have $\sigma_\mathrm{int}(K)$ that peaks at $\sim 0.3$ near the core ($r \sim 0.1R_\mathrm{500c}$). We find lower values, $\sigma_{int}(K) \sim 0.2$, after converting to base-e. Part of these differences may lie in the propagation of uncertainties of halo mass estimates for the X-ray clusters. This can lead to an over-estimation of the intrinsic scatter at fixed radius \citep{Ghirardini2019}. We also find that the minimum in the scatter for the \Flamingo clusters happens at larger radii than for the observed clusters; in \cite{Riva2024}, $r_\mathrm{min} \sim 0.3-0.5\ R_\mathrm{500c}$ for the HIGHMz sample while $r_\mathrm{min} \sim 0.7\  R_\mathrm{500c}$ in our [$14.5-15$] \Flamingo objects. We find a similar pattern when comparing our results to the intrinsic scatter measured for (spectroscopically weighted) temperature profiles in \cite{Rossetti_2024}. This aligns with our prior conclusions that feedback in our simulations has had a stronger effect on the hot gas profiles than in the observed objects.  

\section{Summary and Conclusions}
\label{ch:conclusion}
In this paper, we used the \Flamingo simulation suite (\citealt{flamingo}; \citealt{kugel2023flamingo}) to study a large statistical sample of haloes ranging from groups ($\sim 10^{13}\ \mathrm{M_{\odot}}$) to massive clusters ($\sim 10^{15}\ \mathrm{M_{\odot}}$) up to redshift $z = 1$. We specifically sought to understand the extent to which the dynamical state of these objects, as categorized by three widely used proxies, is reflected in the thermodynamic profiles of the hot gas at different redshifts. The observational proxies we studied were the stellar mass gap between the first and fourth brightest galaxy, $\Mstar$; the X-ray luminosity concentration of the hot gas, $c_\mathrm{x}$ and the X-ray centroid shift parameter, $\left < w \right >$. These were all compared with the underlying halo mass accretion rate, $\Gamma$, which was taken as a direct measure of assembly history. Taking advantage of the \Flamingo runs with alternative feedback models, we determined the role these astrophysical processes play in altering the gas profiles at higher redshift and at lower masses. Our main results and conclusions are summarized as follows:

\begin{itemize}

   \item Dividing our halo sample into mass and redshift bins, we found that the dynamical state proxies are, in general, only weakly correlated with one another for low redshift groups. These correlations are stronger in the higher mass clusters (Fig.\ref{fig:corrs_m8}), where feedback effects are less important and gravitational phenomena dominate. 
 
    \item Our low redshift clusters have $\centshift$ values that are in broad agreement with X-ray observations but their X-ray concentrations are lower. This extends to the case with the weakest AGN feedback model (Fig.\ref{fig:chex_compare}). We relate this to the difficulty with which the simulations have in producing power-law, cool-core clusters with low central entropy and high central X-ray luminosity. The concentrations are even lower in our higher resolution simulation (Fig.\ref{fig:medians}), suggesting that $c_\mathrm{x}$ is not a very robust dynamical state proxy. This is reflected particularly in groups and low-mass clusters, due to their susceptibility to astrophysical feedback processes. 
 
    \item The median values of the other proxies (and the concentrations in the fiducial resolution simulations) are consistent with objects at a fixed mass becoming more disturbed at higher redshift, as expected from the hierarchical growth of structure (Fig.\ref{fig:medians}). These trends are also seen for the different feedback models (Fig.\ref{fig:medians_vars}). We also find that the correlations between each proxy and $\Gamma$ get stronger in the groups at higher redshift (Fig.\ref{fig:corrs_redshift}), where they are dynamically younger on average. Our results suggest the centroid shift to be the most reliable proxy out of those we considered, being reasonably well correlated with $\Gamma$ ($\rho \sim 0.5$) across all our mass and redshift bins, and being the least sensitive to the details of the feedback models. 
    
    \item We used the tails in the distributions of our dynamical state indicators to isolate disturbed and relaxed haloes in each ($M_\mathrm{500c}$, $z$) subset, and measured their average thermodynamic gas properties. Relaxed haloes have density and temperature profiles that are relatively self-similar between $z=0$ and $z=1$, outside of the core. Disturbed haloes, however, display non-self-similar evolution with redshift: at fixed mass, and once the expected self-similar redshift scaling has been factored out, their densities are higher and their temperatures lower at $z=1$ than in their $z=0$ counterparts (Fig. \ref{fig:profiles_dist_rel_gas}). We attribute the enhanced central gas densities in disturbed clusters at high redshift to astrophysical processes, with the gas being able to cool more efficiently. In contrast, the dark matter component evolves similarly in both relaxed and disturbed clusters, with central densities decreasing at higher redshift due to the higher merger activity at early times (Fig. \ref{fig:self_sim_DM}). This behavior is expected, as at fixed mass, disturbed clusters tend to have more of their mass at larger radii due to the presence of secondary objects.

    \item We measured the Pearson correlation coefficient between dynamical state indicator and density at each radius, finding the profile for the dark matter density and $\Gamma$ to have a well defined shape, with the two properties being anti-correlated at small radii and positively correlated at larger radii. This profile is very similar for different masses and redshifts, and is again due to the disturbed objects having more mass at larger radii (Fig. \ref{fig:riva_full}). Similar results are seen for the gas electron density in the $z=0$ clusters (where the signal is strongest) but this is partly washed out on group-scales and at $z=1$, where cooling and feedback effects become more prevalent. From our observable dynamical state probes, $\centshift$ reflects the $\Gamma$ trends more closely than $\Mstar$. We also find a much weaker correlation signal in the clusters with the strongest AGN feedback (Fig. \ref{fig:riva_model}).   

     \item Finally, we studied the evolution of the intrinsic scatter in the profiles with redshift, to understand whether feedback could also affect any underlying trends due to dynamical state (Figs. \ref{fig:scatter_z} and \ref{fig:scatter_model}). Unlike the dark matter density, which shows increased scatter at higher redshift as a result of higher dynamical activity, the scatter reduces with redshift for the gas density profiles, particularly in the groups. We also find the radius of minimum scatter to increase with decreasing halo mass in the case of the gas density profiles but to remain self-similar for the dark matter. This radius is also larger for the strongest feedback model at both low and high redshift, without significantly affecting the amplitude of the scatter, suggesting that this radius could be a useful feedback diagnostic. Comparing the scatter qualitatively with recent low redshift X-ray observations, we find the radius of minimum  scatter to be larger in the simulations. However, a more detailed study taking into account observational effects such as projection will be required to verify this difference, and so we leave this to future work. 
    
\end{itemize}

In summary, our analysis suggests that, although dynamical activity is more common at higher redshift, its impact on the hot gas thermodynamic profiles becomes increasingly difficult to disentangle. At early times, the abundance of smaller haloes ensures that feedback-dominated systems prevail in the sample, enhancing the role of astrophysical processes in shaping gas properties. If confirmed, the reduced diversity in high-redshift gas profiles could simplify how we account for selection effects in deep X-ray and SZ surveys. This is particularly relevant for cluster cosmology, as upcoming surveys such as Euclid and the Simons Observatory will rely more heavily on higher-redshift, lower-mass objects to boost sample sizes and improve cosmological measurements, including constraints on the growth rate of structure and the equation of state of dark energy.

\section*{Acknowledgments}
The research leading to these results has received funding from the European Union’s Horizon 2020 Programme under the AHEAD2020 project (grant agreement n. 871158). This work used the DiRAC@Durham facility managed by the Institute for Computational Cosmology on behalf of the STFC DiRAC HPC Facility (\href{www.dirac.ac.uk}{www.dirac.ac.uk}). The equipment was funded by BEIS capital funding via STFC capital grants ST/K00042X/1, ST/P002293/1, ST/R002371/1 and ST/S002502/1, Durham University and STFC operations grant ST/R000832/1. DiRAC is part of the National e-Infrastructure.
\section*{Data Availability}
The data underlying this article will be shared on reasonable request to the corresponding author.






\bsp	
\label{lastpage}
\end{document}